\documentclass[aps,prd,preprint,groupedaddress]{revtex4-1}

\usepackage{amsmath,amssymb,graphics,epsfig,subfigure}%
\usepackage{color}

\begin{document}


\title{ Localization mechanism of $q-$form field on the brane-world by coupling with gravity }


\author{Yong-Tao Lu }
\email[]{luyt@stu.xidian.edu.cn}
\affiliation{School of Physics, Xidian University, Xi'an 710071, China}
\author{Heng Guo \footnote{Corresponding author}}
\email[]{hguo@xidian.edu.cn}
\affiliation{School of Physics, Xidian University, Xi'an 710071, China}
\author{Chun-E Fu }
\email[]{fuche13@mail.xjtu.edu.cn}
\affiliation{Institute of Theoretical Physics, School of Physics, Xian Jiaotong University,
              Xi'an 710049, People's Republic of China}
\author{Qun Wei }
\email[]{qunwei@xidian.edu.cn}
\affiliation{School of Physics, Xidian University, Xi'an 710071, China}


\date{\today}

\begin{abstract}
It is known that the scalar fields can be trapped on branes of different types, and the $U(1)$ gauge
vector fields can be localized on the thick de Sitter brane, or the thick Minkowski brane via coupling
with gravity. The Kalb-Ramond fields can be localized on the Minkowski brane and the thick de-Sitter
brane, with certain couplings. In this paper, with considering a coupling mechanism between the kinetic
term of the $q-$form fields and the background spacetime, we investigate the localization of Kaluza-Klein
modes for the $q-$form fields in $D-$dimensional spacetime. Concrete $q-$form fields are discussed within
five-dimensional brane models with typical spacetime geometries: Minkowski, de Sitter, and Anti-de Sitter.
In the Minkowski brane case, the zero modes of various $q-$form fields can be localized on the brane. In
the de Sitter brane case, the zero mode of the $U(1)$ gauge vector fields can be localized on the brane.
Lastly in the Anti-de Sitter brane case, the zero mode of the Kalb-Ramond fields can be localized on the
brane. For the massive Kaluza-Klein modes of these $q-$form fields, they could be localized or
quasi-localized on the brane of different types. Besides, subtle and detailed behaviors of the
Kaluza-Klein modes for $q-$form fields are observed: the zero modes could be localized on both sides of
the brane, and the massive modes could be localized or quasi-localized at the brane position.

\end{abstract}



\maketitle


\section{Introduction}

The idea of embedding our universe in higher dimensions has garnered considerable
attention during the last three decades. Exploring the possibility of non-compact
\cite{RubakovPLB1983136,RubakovPLB1983139,SquiresPLB1986286,VisserPLB1985,Randjbar-DaemiPLB1986,rs,RSPRL19994690,Lykken}
or large \cite{AntoniadisPLB1990,ADD,ADDPLB1998257} extra dimensions offers a fresh perspective on
addressing several perplexing phenomena such as the gauge hierarchy \cite{ADD,ADDPLB1998257}, the
dark matter origin and the long-standing cosmological constant problem \cite{RubakovPLB1983136,
RubakovPLB1983139,SquiresPLB1986286,Randjbar-DaemiPLB1986,CosmConst,Kachru2000045021,KehagiasPLB2004133,
StojkovicCHM,Starkman2001231303,ShtanovJCAP2009}. In the Randall-Sundrum (RS) braneworld model
\cite{rs,RSPRL19994690}, the effective four-dimensional (4D) gravity could be recovered in the
case of noncompact extra dimensions, albeit with singularities present at the brane positions.
The majority of thick branes are typically constructed through gravity coupled to bulk scalar
fields \cite{GremmPLB2000,DeWolfePRD2000,CsakiNPB2000,AWangPRD,BazeiaJCAP0402,BazeiaJHEP0405,
BazeiaJHEP0611,AfonsoPLB06634,SlatyerJHEP0704,DzhunushalievGC0713,DzhunushalievPRD0877,BazeiaPRB09671,
ZHZhaoPRD1082,YXLiuEL1090,YXLiuJHEP111106,HGuoPRD1285,HGuoEL1297,VisinelliPRD1897}, and a few
utilize vector fields or spinor fields \cite{DzhunushalievGRG1143,DzhunushalievGRG1244,WJGengPRD1693,
ZQCuiEPJC2383,DzhunushalievEPJC2383}. Moreover, there exist thick branes generated from pure geometry,
devoid of matter field inclusion \cite{Arias02643,BarbosaJHEP0510,BarbosaPRD0673,BarbosaPRD0877,
YXLiuJHEP1010,DzhunushalievJHEP102010,HLiuJHEP1212,JYangPRD1285,YZhongEPJC1676,HGuoPRD23107}.

In braneworld scenarios, a significant and intriguing area of exploration involves the study of
localizing the Kaluza-Klein (KK) modes of various fields \cite{ZHZhaoPRD1082,HGuoPRD1285,
HGuoPRD23107,GrossmanPLB00474,BajcPLB00474,GremmPLB00478,RandjbarPLB00492,OdaPLB00496,KehagiasPLB01508,
OdaPLB01508,IchinosePRD0266,DaviesPRD0877,YXLiuJHEP0808,YXLiuPRD0980,YXLiuPRD0980-2,YXLiuJHEP1002,
YXLiuPRD1184,ArcherJHEP1103,CEFuJHEP1210,HGuoPRD1387,JonesPRD1388,QYXiePRD1388,CostaPRD1387,
ZHZhaoPRD1490,GAlencar14739,ZHZhaoCQG1532,CEFuPRD1693,GAlencar1693,CEFuOLB16757,ZHZhaoJHEP2018,
ZHZhaoJCAP2307,guo2023scalar}. This research delves into the investigation of a bulk massless $q-$form field
on a $p-$brane with codimension-one. It is known that the $0-$form and $1-$form fields represent
the scalar and vector fields, respectively. Furthermore, the usual $2-$form fields correspond to
the Kalb-Ramond (KR) fields, utilized in describing spacetime torsion within Einstein-Cartan theory.
The $p-$brane under consideration possesses $p$ spatial dimensions and is embedded in a $D=p+2$
dimensional bulk spacetime, with one extra dimension orthogonal to the brane.

In Ref. \cite{CEFuJHEP1210}, the authors extensively discussed the localization of a massless
$q-$form field with various $D-$dimensional brane models. More reports can be found in
Refs. \cite{RRLandimEPL2012,CEFuPLB2014,CEFuPRD1693,CEFuOLB16757,GAlencarEPJC2018,RIOliveiraJUniorPLB2022}.
Regarding specific $q-$form fields, the scalar fields can be trapped on branes of different
types \cite{BajcPLB00474}. The $U(1)$ gauge vector fields can be localized in some higher
dimensional cases \cite{OdaPLB00496}, as well as on the thick de Sitter (dS$_4$) brane \cite{YXLiuJCAP0902,
HGuoPRD1387,HerreraEPJC1474}, the Weyl thick brane \cite{YXLiuJHEP0802,YXLiuJHEP0808}, and the
thick Minkowski ($\mathcal{M}_4$) brane with dilaton scalar coupling \cite{YXLiuPRD1285}.
Concerning the KR fields, Refs. \cite{MukhopadhyayaPRL0289,MukhopadhyayaPRD0470} identified such
field as representing bulk torsion, determining its effect to be notably weaker than curvature.
This characteristic makes their detection challenging on the RS thin brane. However, when considered
as the perturbation to the bulk, the KR fields can be localized on the thick $\mathcal{M}_4$
brane with dilaton scalar coupling \cite{TahimPRD0979,CruzEL0988,CEFuPRD1184,YXLiuPRD1285,CruzEPJC1373}
or background scalar coupling \cite{YZDuPRD1388}, and on the thick dS$_4$ brane with background
scalar coupling \cite{Yang1911.11438}. There are less reports on the KR fields within the Anti-de Sitter
(AdS$_4$) brane model. For comprehensive reviews, see Refs. \cite{YXLiu1707.08541,AhluwaliaPR967}.

In this paper, we explore a coupling mechanism between the kinetic term of the $q-$form fields and the
background spacetime. The coupling mechanism is very significative for the relativistic particles.
Without introducing additional coupling terms, the action for a massless $q-$form field incorporates
a multiplier factor $F(R)$, which is a function of the scalar curvature of the bulk spacetime. For an
elementary field, its kinetic energy is far larger than its rest energy in magnitude. It is both
rational and physically motivated to consider this coupling mechanism. Then, we further investigate
the localization of the KK modes for the $q-$form fields, and demonstrate:
\emph{\begin{itemize}
  \item (a) A general analysis on the localization mechanism of the $q-$form fields coupled with the gravity
            in $D-$dimesional spacetime with codimension-one.
  \item (b) Three kinds of brane cases with typical geometries: Minkowski, de Sitter, and Anti-de Sitter, are
            examined.
  \item (c) Subtle and novel behaviors of the $q-$form fields at finite positions of the extra dimension.
\end{itemize}}
We find that various $q-$form fields can be localized on the thick branes of different types, excluding
the KR fields on thick dS$_4$ branes or the $U(1)$ gauge vector fields on thick AdS$_4$ brane. Furthermore,
at finite positions of the extra dimension, the zero modes of the $q-$form fields could be localized on
both sides of the brane, with their probability densities exhibiting a split along the extra dimension.

The paper is structured as follows: We introduce the coupling mechanism between the kinetic term of the $q-$form
field and the background spacetime in Sec. \ref{sec2}. Then, Sec. \ref{sec3} and \ref{sec4} delve into the
investigation on the localization of the zero mode and the massive modes, for a massless $q-$form field in
$D-$dimensional spacetime. Moving on to Sec. \ref{sec5}, we discuss the localization of concrete $q-$form
fields$-$specifically, the scalar fields, the $U(1)$ gauge vector fields, and the KR fields. Three kinds of
brane cases are considered: $\mathcal{M}_4$ brane, dS$_4$ brane, and AdS$_4$ brane. Finally, the conclusions
are presented in Sec. \ref{Cons}.



\section{The {  Method} }\label{sec2}

The line element of the $D-$dimensional bulk spacetime with $D=p+2$ is assumed as
\begin{eqnarray}\label{metric}
 ds^2=g_{MN}dx^Mdx^N=e^{2A(y)}\hat g_{\mu\nu}dx^{\mu}dx^{\nu}+dy^2,
\end{eqnarray}
where $e^{2A(y)}$ is the warp factor, $\hat g_{\mu\nu}$ is the metric on the $p-$brane and $y$ denotes the extra dimension. The capital
Latin letters $M,N,...=0,1,2,...p+1$ and the Greek letters $\mu,\nu,...=0,1,2,...p$ are used to represent the bulk and brane indices,
respectively. From this metric, the scalar curvature of the bulk spacetime can be given by
\begin{eqnarray}\label{curvature}
 R=e^{-2A}\hat R-2(p+1)A''-(p+1)(p+2) A'^2,
\end{eqnarray}
where $\hat R$ is the scalar curvature of the brane spacetime. The $D-$dimensional spacetime with asymptotically constant
scalar curvature means
\begin{eqnarray} \label{ItemSptAsymD}
    R\rightarrow
\left\{
  \begin{array}{ll}
    \text C_{R_1}      & \hspace{0.5cm} \text{asymptotically dS}   \\
    0                  & \hspace{0.5cm} \text{asymptotically flat} \\
    -\text C_{R_2}     & \hspace{0.5cm} \text{asymptotically AdS}
  \end{array}
\right.
\end{eqnarray}
when far away from the brane, where $\text C_{R_1}$ and $\text C_{R_2}$ are positive constants. The brane model
considered here has no singularity for the scalar curvature, which is regular.

Considering a $D-$dimensional massless antisymmetric $q-$form field $X_{M_1M_2...M_{q+1}}$, we treat
this $q-$form field as a small perturbation around the background spacetime, and assume its action as
\begin{eqnarray}\label{action}
 S_q=\int d^Dx\sqrt{-g}F(R)Y_{M_1M_2...M_{q+1}}Y^{M_1M_2...M_{q+1}},
\end{eqnarray}
where $Y_{M_1M_2...M_{q+1}}$ is the field strength, defined as $Y_{M_1M_2...M_{q+1}}=\partial_{[M_1}X_{M_2...M_{q+1}]}$.
The factor $F(R)$ is a function of the scalar curvature of the bulk, and it stands for the coupling between
the kinetic term of the $q-$form field and the background spacetime. Since the $(D-1)-$form and the higher
form fields play no role in the brane, our analysis focuses solely on cases where the index $q$ ranges from
$0$ to $p$.

The function $F(R)$ should adhere to the following rules:
\begin{enumerate}
\item The scalar curvature $R$ of the bulk, and the coupling factor $F(R)$ should be nonsingular.
\item If the scalar curvature $R \to 0$, the bulk spacetime becomes flat.
The coupling should turn into the minimal coupling with $F(R)\rightarrow1$, in order that
the action (\ref{action}) returns to the one of a free $q-$form field:
\begin{equation} \label{action0}
S_q=\int d^Dx\sqrt{-g}Y_{M_1M_2...M_{q+1}}Y^{M_1M_2...M_{q+1}}.
\end{equation}
\item { The function $F(R)$ should satisfy the positivity condition
\begin{equation}
  F(R) >0 \label{positivity}
\end{equation}
 to preserve the canonical form of $(D-1)-$dimensional action}.
\end{enumerate}

This coupling has also been discussed in Ref. \cite{ZHZhaoJHEP2018}, where three forms for the function
$F(R)$ were proposed: two polynomial forms and one exponential form. However, an exponential function
can always be expanded into a polynomial. Additionally, the dilaton scalar field $\pi$, which exhibits
the same even-parity as the scalar curvature, appears in the exponential form $e^{\xi\pi}$ when coupled
with the kinetic term \cite{KehagiasPLB01508,CEFuPRD1184}. Considering the similarity between the function
$F(R)$ and the dilaton scalar $\pi$, we will suggest the former in exponential form.

\section{Localization of the Zero Mode}\label{sec3}

To aid in analyzing the localization of the KK modes for the $q-$form fields, we will
transition to the conformal coordinate $z$ through the coordinate transformation:
\begin{eqnarray} \label{coordinate trans}
 \left\{
   \begin{array}{ll}
     & dz=e^{-A(y)}dy \\
     & z=\int e^{-A(y)}dy
   \end{array}
 \right.
\end{eqnarray}
with the boundary condition $z(y=0)=0$. Then, the line element (\ref{metric}) can be expressed in
terms of coordinate $z$:
\begin{eqnarray} \label{metricz}
ds^2=e^{2A}(\hat g_{\mu\nu}dx^{\mu}dx^{\nu}+dz^2).
\end{eqnarray}

For a bulk $q-$form field, we make a general KK decomposition:
\begin{eqnarray} \label{DecomDq}
  X_{M_1M_2...M_q}(x^{\lambda},z)=\sum_n\hat X^{(n)}_{M_1M_2...M_q}(x^{\lambda})U^{(n)}(z),
\end{eqnarray}
where $U^{(n)}(z)$ are the KK modes for the $q-$form field. The action of the
massless $q-$form field (\ref{action}) remains invariant under the following gauge
transformation with an arbitrary antisymmetric tensor $\Lambda_{M_2M_3...M_q}$
\cite{MukhopadhyayaPRD07121501}:
\begin{eqnarray} \label{GT1}
  X_{M_1M_2...M_q}(x^{\lambda},z)\rightarrow \tilde X_{M_1M_2...M_q}(x^{\lambda},z)
    =X_{M_1M_2...M_q}(x^{\lambda},z)+\partial_{[M_1}\Lambda_{M_2...M_q]},
\end{eqnarray}
or
\begin{eqnarray}
  X_{\mu_1M_2...M_q}(x^{\lambda},z)&\rightarrow& \tilde X_{\mu_1M_2...M_q}(x^{\lambda},z)
    =X_{\mu_1M_2...M_q}(x^{\lambda},z)+\partial_{[\mu_1}\Lambda_{M_2...M_q]},            \label{GT2}    \\
  X_{zM_2...M_q}(x^{\lambda},z)&\rightarrow& \tilde X_{zM_2...M_q}(x^{\lambda},z)
    =X_{zM_2...M_q}(x^{\lambda},z)+\partial_{[z}\Lambda_{M_2...M_q]}.                    \label{GT3}
\end{eqnarray}
Subsequently, we will assess whether the component $X_{zM_2...M_q}(x^{\lambda},z)$ can be
nullified by the mentioned gauge transformation.

Given the presence of infinite extra dimension in our braneworld scenario, there exist no
constraints on the tensors $X_{M_1M_2...M_q}(x^{\lambda},z)$ or
$\Lambda_{M_2M_3...M_q}(x^{\lambda},z)$. Utilizing both the transformation (\ref{GT3}) and
the KK decomposition (\ref{DecomDq}), we derive:
\begin{eqnarray}  \label{GT4}
  X_{zM_2...M_q}(x^{\lambda},z)&\rightarrow& \tilde X_{zM_2...M_q}(x^{\lambda},z)
    =\sum_n\hat X^{(n)}_{zM_2...M_q}(x^{\lambda})U^{(n)}(z)+\partial_{[z}\Lambda_{M_2...M_q]}.
\end{eqnarray}
Thus, if we choose the tensor $\Lambda_{M_2M_3...M_q}(x^{\lambda},z)$ as
\begin{eqnarray}  \label{GT5}
  \Lambda_{M_2M_3...M_q}(x^{\lambda},z)=-\sum_n\hat X^{(n)}_{zM_2...M_q}(x^{\lambda})
     \int U^{(n)}(z) dz,
\end{eqnarray}
the extra-dimensional component $\tilde X_{zM_2...M_q}$ will vanish:
\begin{eqnarray}  \label{GT6}
  \tilde X_{zM_2...M_q}(x^{\lambda},z)=0,
\end{eqnarray}
which is just the gauge choice we made.

Hence, we set $X_{zM_2...M_q}(x^{\lambda},z)=0$ by using this gauge freedom. With this gauge, the
non-vanishing component of $X_{M_1M_2...M_q}(x^{\lambda},z)$ is $X_{\mu_1\mu_2...\mu_q}(x^{\lambda},z)$,
and then, the KK decomposition (\ref{DecomDq}) becomes
\begin{eqnarray} \label{Decomq}
  X_{\mu_1\mu_2...\mu_q}(x^{\lambda},z)=\sum_n\hat X^{(n)}_{\mu_1\mu_2...\mu_q}(x^{\lambda})U^{(n)}(z).
\end{eqnarray}
Therefore, the field strength becomes
\begin{eqnarray}
Y_{\mu_1\mu_2...\mu_{q+1}}(x^{\lambda},z)&=&\sum_n\hat Y^{(n)}_{\mu_1\mu_2...\mu_{q+1}}(x^{\lambda})
  U^{(n)}(z),                                              \label{field strength01}           \\
Y_{\mu_1\mu_2...\mu_q z}(x^{\lambda},z)&=&\sum_n\frac{1}{q+1}
  \hat X^{(n)}_{\mu_1\mu_2...\mu_q}(x^{\lambda})U^{(n)}{'}(z),    \label{field strength02}
\end{eqnarray}
where $\hat Y^{(n)}_{\mu_1\mu_2...\mu_{q+1}}(x^{\lambda})=\partial_{[\mu_1}\hat X_{\mu_2...\mu_{q+1}]}(x^{\lambda})$
is the field strength on the $p-$brane, and the prime denotes the derivative with respect
to coordinate $z$ here. In terms of Eqs. (\ref{field strength01}) and (\ref{field strength02}),
the action (\ref{action}) can be simplified to
\begin{eqnarray} \label{action4D}
S_q&=&\sum_n\int dze^{(p-2q)A}F(R)(U^{(n)})^2\int d^{D-1}x\sqrt{-\hat g}
      \bigg(\hat Y^{(n)}_{\mu_1\mu_2...\mu_{q+1}}\hat Y^{(n)\mu_1\mu_2...\mu_{q+1}}            \nonumber      \\
   & &+\frac{1}{q+1}m^2_n\hat X^{(n)}_{\mu_1\mu_2...\mu_q}\hat X^{(n)\mu_1\mu_2...\mu_q}\bigg),
\end{eqnarray}
where $m_n$ are the masses of the KK modes, and $U^{(n)}$ satisfy the equation
\begin{eqnarray} \label{EquKKmode0}
U^{(n)}{''}+\bigg((p-2q)A'+\frac{F'(R)}{F(R)}\bigg)U^{(n)}{'}=-m^2_nU^{(n)}
\end{eqnarray}
with the boundary conditions either $U^{(n)}{'}(\pm\infty)=0$ or $U^{(n)}(\pm\infty)=0$ \cite{Gherghetta1008.2570}.
The localization of the $q-$form field requires
\begin{eqnarray} \label{LocReq}
\int e^{(p-2q)A}F(R)(U^{(n)})^2dz=1.
\end{eqnarray}

For the zero mode, $m_0^2=0$, Eq. (\ref{EquKKmode0}) becomes
\begin{eqnarray} \label{Equzeromode0}
U_0{''}+\bigg((p-2q)A'+\frac{F'(R)}{F(R)}\bigg)U_0{'}=0.
\end{eqnarray}
By setting $\gamma'=(p-2q)A'+F'(R)/F(R)$, the above equation can be expressed as
\begin{eqnarray} \label{Equzeromode1}
U_0{''}+\gamma'U_0{'}=0.
\end{eqnarray}
From this expression, we can derive the following general solution for the zero mode:
\begin{eqnarray} \label{Genezeromode}
U_0=\text{c}_0+\text{c}_1\int e^{-\gamma}dz,
\end{eqnarray}
where the parameters $\text{c}_0$ and $\text{c}_1$ are integral constants. The models presented
below display $\mathbb{Z}_2$ symmetry along the extra dimension, making the second
term in Eq. (\ref{Genezeromode}) odd. The Dirichlet boundary conditions $U^{(n)}(\pm\infty)=0$
will result in $\text{c}_0=0$ and $\text{c}_1=0$, while the Neumann boundary conditions
$U^{(n)}{'}(\pm\infty)=0$ only yield $\text{c}_1=0$. Consequently, the zero mode solution
is
\begin{eqnarray} \label{Soluzeromode}
U_0=\text{c}_0.
\end{eqnarray}
According to Eq. (\ref{LocReq}), the localization of the zero mode requires
\begin{eqnarray} \label{Norzeromode}
   & &\int e^{(p-2q)A}F(R)U_0^2dz                 \nonumber      \\
   &=&\text{c}_0^2\int e^{(p-2q)A}F(R)dz=1.
\end{eqnarray}
With the coordinate transformation (\ref{coordinate trans}), this condition becomes 
\begin{eqnarray} \label{IntNorzeromode}
   \text{c}_0^2\int e^{(p-2q)A}F(R)dz=\text{c}_0^2\int e^{(p-2q-1)A}F(R)dy=1.
\end{eqnarray}
{ 
So the normalization of the zero mode requires 
\begin{eqnarray} \label{Cond1IntNorzm}
   F(R(y\rightarrow\pm\infty))\propto y^{-l}e^{-(p-2q-1)A},
\end{eqnarray}
where $l>1$. At this point, the normalization condition (\ref{IntNorzeromode}) is equivalent to
the following one:
\begin{eqnarray} \label{Cond2IntNorzm}
   \int^{+\infty}_{y_0}y^{-l}dy=\frac{1}{l-1}\frac{1}{y_0^{l-1}}<\infty
\end{eqnarray}
with constant $y_0\gg0$. It can be seen that the normalization condition (\ref{IntNorzeromode}) is satisfied. }

Besides, for the condition (\ref{IntNorzeromode}), the localization of the zero mode for the $q-$form field
is model-dependent. The further analysis will be conducted with concrete brane models. A specific condition
is that if there is an asymptotically flat brane model, the coupling function $F(R)$ converges to a
constant when far away from the brane. Then, introducing the function $F(R)$ cannot influence the
localization result of the zero mode.


\section{Localization of Massive modes}\label{sec4}

Considering the massive KK modes for the $q-$form field, we will focus on the effective
potentials of the KK modes in the corresponding Schr\"{o}dinger-like equations. Beginning with
the metric (\ref{metricz}), the equations of motion can be derived from the action (\ref{action}):
\begin{eqnarray}
\partial_{\mu_1}(\sqrt{-g}F(R)Y^{\mu_1\mu_2...\mu_{q+1}})+
  \partial_z(\sqrt{-g}F(R)Y^{z\mu_2...\mu_{q+1}})&=&0,    \label{EoM1}           \\
\partial_{\mu_1}(\sqrt{-g}Y^{\mu_1\mu_2...\mu_qz})&=&0.   \label{EoM2}
\end{eqnarray}

Then, with the KK decomposition for the $q-$form field $X_{\mu_1\mu_2...\mu_q}$:
\begin{eqnarray} \label{Decom}
X_{\mu_1\mu_2...\mu_q}=\sum_n\hat{X}^{(n)}_{\mu_1\mu_2...\mu_q}(x^{\lambda})
  \tilde U^{(n)}(z)e^{\frac{2q-p}{2}A}(F(R))^{-\frac12},
\end{eqnarray}
where $\tilde U^{(n)}(z)=U^{(n)}(z)e^{-\frac{2q-p}{2}A}(F(R))^{\frac12}$, the field strength
becomes
\begin{eqnarray}
Y_{\mu_1\mu_2...\mu_{q+1}}(x^{\lambda},z)&=&\sum_n\hat Y^{(n)}_{\mu_1\mu_2...\mu_{q+1}}(x^{\lambda})
  \tilde U^{(n)}e^{\frac{2q-p}{2}A}(F(R))^{-\frac12},    \label{field strength1}                        \\
Y_{\mu_1\mu_2...\mu_q z}(x^{\lambda},z)&=&\sum_n\frac{1}{q+1}
  \hat X^{(n)}_{\mu_1\mu_2...\mu_q}(x^{\lambda})\bigg(\tilde U^{(n)}{'}(F(R))^{-\frac12}    \nonumber          \\
 & &-\frac12\tilde U^{(n)}(F(R))^{-\frac32}F'(R)
  +\frac{2q-p}{2}A'\tilde U^{(n)}(F(R))^{-\frac12}\bigg) e^{\frac{2q-p}{2}A}.    \label{field strength2}
\end{eqnarray}

By substituting the relations (\ref{field strength1}) and (\ref{field strength2}) into the
equation of motion (\ref{EoM1}), we can obtain the following Schr\"{o}dinger-like equation for
the KK modes:
\begin{eqnarray} \label{Sch-equa}
[-\partial^2_z+V(z)]\tilde U^{(n)}(z)=m^2_n\tilde U^{(n)}(z),
\end{eqnarray}
where the effective potential $V(z)$ is
\begin{eqnarray} \label{eff-potential}
V(z)=\frac{p-2q}{2}A''+\frac{(p-2q)^2}{4}A'^2+\frac{p-2q}{2}\frac{A'F'(R)}{F(R)}+
     \frac{F''(R)}{2F(R)}-\frac{F'^2(R)}{4F^2(R)}.
\end{eqnarray}

Furthermore, requiring the orthonormality condition
\begin{eqnarray} \label{ortho}
\int \tilde U^{(m)}\tilde U^{(n)}dz=\delta_{mn},
\end{eqnarray}
we can derive the effective action of the $q-$form field on the brane as follows
\begin{eqnarray} \label{eff-action}
S_{\text{eff}}&=&\sum_n\int d^{D-1}x\sqrt{-\hat g}\bigg(\hat Y^{(n)}_{\mu_1\mu_2...\mu_{q+1}}
  \hat Y^{(n)\mu_1\mu_2...\mu_{q+1}}                            \nonumber  \\
  & &+\frac{1}{q+1}m^2_n
  \hat X^{(n)}_{\mu_1\mu_2...\mu_q}\hat X^{(n)\mu_1\mu_2...\mu_q}\bigg).
\end{eqnarray}
The KK modes of the $q-$form field can be localized on the thick brane when the orthonormality condition
(\ref{ortho}) is satisfied. Therefore, the solutions of the Schr\"{o}dinger-like equation (\ref{Sch-equa})
are shaped by the effective potential (\ref{eff-potential}), and the localization result of the KK modes
is determined ultimately by both the warp factor and the coupling function $F(R)$.

The Schr\"{o}dinger-like equation (\ref{Sch-equa}) can further be recast to
\begin{eqnarray} \label{diracEqua}
 QQ^{\dagger}\tilde U^{(n)}(z)=m^2_n\tilde U^{(n)}(z)
\end{eqnarray}
with
\begin{eqnarray} \label{DerGammaz}
   Q  &=& \partial_z+\bigg(\frac{p-2q}{2}A'+\frac{F'(R)}{2F(R)}\bigg),       \\
   Q^{\dagger} &=& -\partial_z+\bigg(\frac{p-2q}{2}A'+\frac{F'(R)}{2F(R)}\bigg).
\end{eqnarray}
Equation (\ref{diracEqua}) implies that there is no tachyonic mode with $m^2_n<0$
in the spectrum of the KK modes \cite{BazeiaJCAP0402}. In addition, by setting
$m^2_0=0$, we can obtain the solution for the zero mode:
\begin{eqnarray} \label{zero mode}
 \tilde U_0(z)=Ne^{\frac{p-2q}{2}A}(F(R))^{\frac12}
\end{eqnarray}
with $N$ the normalization constant. Based on the coordinate transformation (\ref{coordinate trans}), this
zero mode solution can be re-expressed in terms of coordinate $y$ as
\begin{eqnarray} \label{yzero mode}
 \tilde U_0(z(y))=Ne^{\frac{p-2q}{2}A(y)}(F(R))^{\frac12}.
\end{eqnarray}
Furthermore, the localization of the zero mode requires:
\begin{eqnarray} \label{CondLocaZM}
 \int \tilde U_0^2 dz = N^2\int e^{(p-2q)A}F(R)dz=1,
\end{eqnarray}
which is equivalent to the localization condition (\ref{Norzeromode}).

\section{Localization of Various $q-$form Fields}\label{sec5}

In this section, we will investigate the localization of various $q-$form fields in 5D
spacetime. Specifically, we consider the case where the index $p=3$. In this context,
the capital Latin letters $M,N,...=0,1,2,3,5$ and Greek letters $\mu,\nu,...=0,1,2,3$
are used to stand for the bulk and brane indices, respectively. Three types of brane
cases are examined: the $\mathcal{M}_4$ brane, the dS$_4$ brane, and the AdS$_4$ brane.
It is implicitly assumed that the various $q-$form fields considered here make negligible
contributions to the bulk energy. Therefore, the brane solutions presented below remain
valid even in the presence of bulk matter.

\subsection{Minkowski brane} \label{flatbrane}

In the $\mathcal{M}_4$ brane case, the brane spacetime is flat, and the line element is
\begin{eqnarray} \label{metricMin}
ds^2=e^{2A}(\eta_{\mu\nu}dx^{\mu}dx^{\nu}+dz^2),
\end{eqnarray}
where $\eta_{\mu\nu}$ is the metric of the flat brane. For this case, we consider the RS-II thick brane
model \cite{GremmPLB2000,CEFuJHEP1210}:
\begin{eqnarray}
    A(y) &=& -\frac{b}{2}\ln\big(\cosh(cy)\big),         \label{Warped factor}        \\
    \varphi(y) &=& \sqrt{6b}\arctan\big(\tanh(\frac{cy}{2})\big),
\end{eqnarray}
where $b$ and $c$ are positive constants. For this brane model, the scalar curvature of the bulk is
\begin{eqnarray}\label{curvatureMin}
 R=bc^2\big[-5b+(5b+4)\text{sech}^2(cy)\big].
\end{eqnarray}
At the origin and infinity, it becomes
\begin{eqnarray}\label{AsycurvatureMin}
 R(y=0) &=&  4bc^2,                                                       \\
 R(y\rightarrow +\infty) &\rightarrow& -5b^2c^2.
\end{eqnarray}
Since the brane model (\ref{Warped factor}) holds $\mathbb{Z}_2$ symmetry, we only consider the
asymptotic behavior of $A(y)$ as $y\rightarrow+\infty$, and the same applies to all other quantities
subsequently presented. Besides, we can see that the 5D bulk is asymptotically AdS.

{ For this brane model, introducing the coupling function $F(R)$ should facilitate the
localization of the $q-$form fields. As the scalar curvature $R$ decreases monotonously when $y$
varies from $0$ to $\pm\infty$, based on the three rules of $F(R)$, we can have a simple form:
\begin{eqnarray} \label{FR5-0}
 F(R)=1+\frac{R}{5b^2c^2}.
\end{eqnarray}
Then, the zero mode (\ref{yzero mode}) can be expressed as
\begin{eqnarray} \label{ZMMin-1}
     \tilde U_0(z(y))=N\sqrt{\frac{5b+4}{5b}}(\text{sech}(cy))^{\frac{b}{2}(3-2q)+1}.
\end{eqnarray}
Based on the coordinate translation (\ref{coordinate trans}), the localization condition for the zero
mode (\ref{CondLocaZM}) becomes
\begin{eqnarray} \label{ExprCondLocaZM-0}
 \int \tilde U_0^2 dz &=& N^2\int e^{(p-2q)A}F(R)dz              \nonumber   \\
                      &=& N^2\int e^{(p-2q-1)A(y)}F(R)dy         \nonumber   \\
                      &=& N^2\int \frac{5b+4}{5b}(\text{sech}(cy))^{b-bq+2} dy       \nonumber   \\
                      &=& 1.
\end{eqnarray}
Therefore, the normalization of the zero mode (\ref{ZMMin-1}) requires 
\begin{eqnarray} \label{CondLocaZM-1}
   b-bq+2>0,
\end{eqnarray}
and we can see that the zero modes of both the $q=0$ scalar field and the $q=1$ $U(1)$ gauge vector field can 
always be localized. However, the localization of the zero mode for the $q=2$ KR field requires that the warp 
factor parameter $b<2$.

Considering the massive modes of the $q-$form field, we shall conduct analysis on the Schr\"{o}dinger-like equation
(\ref{Sch-equa}). In the context of the brane model (\ref{Warped factor}), there is no analytical expression for
the effective potential $V(z)$ (\ref{eff-potential}) with respect to coordinate $z$. Thus, utilizing the coordinate
transformation (\ref{coordinate trans}), we re-express it concerning coordinate $y$:
\begin{eqnarray} \label{yeff-potentialMin}
V(z(y))&=& \frac{(3-2q)(5-2q)}{4}A'^2e^{2A}+\frac{3-2q}{2}A''e^{2A}         \nonumber         \\
       & & +\frac{F''(R)e^{2A}}{2F(R)}+(2-q)\frac{A'F'(R)e^{2A}}{F(R)}-\frac{F'^2(R)e^{2A}}{4F^2(R)},
\end{eqnarray}
where the prime stands for the derivative with respect to coordinate $y$ here. Substituting Eqs. (\ref{Warped factor})
and (\ref{FR5-0}) into this expression, we can get
\begin{eqnarray} \label{Expryeff-potentialMin-1}
    V(z(y))=\frac{[b(2q-3)-4]c^2}{16}[4+(2bq-5b-4)(\sinh(cy))^2](\text{sech}(cy))^{b+2}.
\end{eqnarray}
It is clear that since $b>0$, $V(z(y\rightarrow+\infty))\rightarrow0$, so this effective potential takes the 
shape of a volcano. Therefore, when introducing the function $F(R)$ (\ref{FR5-0}), the zero mode of various 
$q-$form fields can be localized on the brane, in some cases with constraints on the parameter $b$. However, 
there are no localized massive KK modes of any $q-$form field on the brane.

Based on the form (\ref{FR5-0}), we can further assume the function $F(R)$ as a polynomial:
\begin{eqnarray} \label{FR5-1}
 F(R)=\frac1n\sum_{i=1}^{n}\bigg(1+\frac{R}{5b^2c^2}\bigg)^i
\end{eqnarray}
with integer $n>0$. This function takes the higher order term of Eq. (\ref{FR5-0}) 
into consideration, and could lead to the similar localization results.

Making more steps ahead, as an exponential function can always be expanded into a polynomial, we suggest 
another form of function $F(R)$ as \cite{ZHZhaoJHEP2018,guo2023scalar}
\begin{eqnarray} \label{FR5}
 F(R)=e^{t_1[1-((1+\frac{R}{5b^2c^2})^2)^{-t_2/2}]},
\end{eqnarray}
where $t_1$ and $t_2$ are coupling parameters with $t_1>0$. In this case,
firstly, the zero mode (\ref{yzero mode}) can be expressed as
\begin{eqnarray} \label{ZMMin}
     \tilde U_0(z(y))=N(\cosh(cy))^{\frac{2q-3}{4}b} \sqrt{e^{t_1\big[1-
              (\frac{5b}{5b+4})^{t_2}(\text{sech}(cy))^{-2t_2}\big]}}.
\end{eqnarray}
Furthermore, the localization condition for the zero
mode (\ref{CondLocaZM}) becomes
\begin{eqnarray} \label{ExprCondLocaZM}
 \int \tilde U_0^2 dz = N^2\int e^{t_1\big[1-(\frac{5b}{5b+4})^{t_2}(\text{sech}(cy))^{-2t_2}\big]}(\cosh(cy))^{b(q-1)}dy=1.
\end{eqnarray}
For this integral, we can obtain the asymptotical solution of its integrand as $y\rightarrow +\infty$:
\begin{eqnarray} \label{AsymExprCondLocaZM}
   N^2e^{(p-2q-1)A(y)}F(R)\rightarrow
          N^22^{-b(q-1)}e^{t_1\big[1-(\frac{5b}{20b+16})^{t_2}e^{2t_2cy}\big]}e^{b(q-1)cy}.
\end{eqnarray}
It can be seen that the realization of the localization condition (\ref{ExprCondLocaZM}) is determined by
both the parameter $t_2$ and the index $q$, with no restraint on parameter $b$. Specifically, the $q=0$
scalar field can always be localized on the brane, and localization of other $q-$form fields requires the
parameter $t_2>0$.

Additionally, at the origin, the zero mode (\ref{ZMMin}) becomes
\begin{eqnarray} \label{OriZMMin}
     \tilde U_0(z(y=0))=N\sqrt{e^{t_1-t_1(\frac{5b}{5b+4})^{t_2}}}.
\end{eqnarray}
Thus, if parameter $t_2<0$, the zero mode will be suppressed to zero at the origin of the extra dimension.


Turning to the massive KK modes, with substituting Eqs. (\ref{Warped factor}) and (\ref{FR5}) into the
expression (\ref{yeff-potentialMin}), we can get
\begin{eqnarray} \label{Expryeff-potentialMin}
\hspace{-0.5cm}V(z(y))&=& \frac{c^2}{16}(\text{sech}(cy))^{b+2}\bigg(16\times25^{t_2}t_1^2t_2^2
           \big(\frac{(5b+4)^2\text{sech}^4(cy)}{b^2}\big)^{-t_2}\sinh^2(cy)       \nonumber      \\
       & & +b(2q-3)[4+b(2q-5)\sinh^2(cy)]                                          \nonumber      \\
       & & -16\times5^{t_2}t_1t_2(\frac{(5b+4)^2\text{sech}^4(cy)}{b^2})^{-\frac{t_2}{2}}
           \big(1+[b(q-2)+2t_2]\sinh^2(cy)\big) \bigg).
\end{eqnarray}
For this effective potential, we can further determine its asymptotical behaviors as $y\rightarrow +\infty$:
\begin{eqnarray} \label{AsymyEffPotMin}
V(z(y\rightarrow +\infty))=
        \left\{
          \begin{array}{ll}
            +\infty    &\hspace{0.5cm} t_2>b/4,    \\
            \text{C}   &\hspace{0.5cm} t_2=b/4,    \\
            0          &\hspace{0.5cm} t_2<b/4
          \end{array}
        \right.
\end{eqnarray}
with the positive limit
\begin{eqnarray} \label{LmtEffPotMin}
 \text{C} = \frac{1}{16}b^2c^2t_1^2\bigg(\frac{5b}{5b+4} \bigg)^{b/2}.
\end{eqnarray}
This effective potential (\ref{Expryeff-potentialMin}) could exhibit more abundant forms.

From the expression (\ref{AsymyEffPotMin}), it can be seen that the condition $t_2=b/4$ means a finite number
of the localized massive modes, when $t_2>b/4$, there will be infinite number of localized massive modes, and
$t_2<b/4$ corresponds to no localized massive mode on the brane.

Except for the localized zero mode, the coupling function $F(R)$ (\ref{FR5}) could also give rise to localized 
massive modes. So, we will focus solely on this form of $F(R)$ in the following discussion on concrete $q-$form 
fields.
}

\subsubsection{Scalar fields}   \label{flatScalar}

The first $q-$form field discussed here is the scalar field with index $q=0$. This field describes the
particles with spin$-0$, and is classified into real scalar field and complex scalar field (denoting
the charged particles) in quantum field theories. In the Standard Model, the Higgs field, which gives
mass to massive elementary particles, belongs to the category of scalar field.

We consider a 5D massless real scalar field $\Phi(x^{\mu},z)$ and assume the action for it as
\begin{eqnarray} \label{ActionSca}
  S_0=-\frac12\int d^5x\sqrt{-g}F(R)\partial_M\Phi\partial^M\Phi.
\end{eqnarray}
With the flat metric (\ref{metricMin}), the equation of motion can be derived from the action as
\begin{eqnarray} \label{EoSSca}
  \partial_{\mu}(\sqrt{-g}F(R)\partial^{\mu}\Phi)+\partial_z(\sqrt{-g}F(R)\partial^z\Phi)=0.
\end{eqnarray}
By using the KK decomposition
\begin{eqnarray} \label{DecomSca}
  \Phi(x^{\mu},z)=\sum_n\phi_n(x^{\mu})\chi_n(z)e^{-\frac32A}(F(R))^{-\frac12}
\end{eqnarray}
and demanding that $\phi_n$ satisfy the 4D massive Klein-Gordon equation:
\begin{eqnarray} \label{KGequation}
  \bigg[\frac{1}{\sqrt{-\hat g}}\partial_{\mu}(\sqrt{-\hat g}\hat{\eta}^{\mu\nu}\partial_{\nu})-m^2_n\bigg]\phi_n(x)=0,
\end{eqnarray}
where $m_n$ are the masses of the scalar KK modes, we can obtain the Schr\"{o}dinger-like equation
\begin{eqnarray} \label{SchroSca}
  \big[-\partial_z^2+V_0(z)\big]\chi_n(z)=m_n^2\chi_n(z)
\end{eqnarray}
with the effective potential
\begin{eqnarray} \label{effVSca}
  V_0(z)=\frac32A''+\frac{9}{4}A'^2+\frac32\frac{A'F'(R)}{F(R)}+\frac{F''(R)}{2F(R)}-\frac{F'^2(R)}{4F^2(R)}.
\end{eqnarray}

By requiring the orthonormality conditions for the KK modes
\begin{eqnarray} \label{orthoSca}
  \int \chi_m(z)\chi_n(z)dz=\delta_{mn},
\end{eqnarray}
the 5D action (\ref{ActionSca}) can be reduced into the 4D effective one:
\begin{eqnarray} \label{effActSca}
  S_0=-\frac12\sum_n\int d^4x\sqrt{-\hat g}(\partial_{\mu}\phi_n\partial^{\mu}\phi^n+m^2_n\phi^2_n),
\end{eqnarray}
when integrated over the extra dimension. The localization of the KK modes need the orthonormality
condition (\ref{orthoSca}) to be satisfied.

At this stage, when we consider the case of minimal coupling with coupling function $F(R)\equiv1$, the
effective potential $V_0(z)$ becomes:
\begin{eqnarray} \label{effVScaMC}
  V_0(z)=\frac32A''+\frac{9}{4}A'^2.
\end{eqnarray}
Based on the RSII-like model (\ref{Warped factor}), this potential $V_0(z)$ (\ref{effVScaMC}) has the
volcano shape, allowing only the scalar zero mode to be localized on the brane, which is the same with
the thin brane case \cite{CEFuJHEP1210}.

Then, the localization of scalar field will be investigated with considering the coupling function $F(R)$
in the following. Moving forward with the Schr\"{o}dinger-like equation (\ref{SchroSca}), by setting
$m_0^2=0$, we can obtain the zero mode solution
\begin{eqnarray} \label{zeromodeSca}
  \chi_0(z)=N_0e^{\frac32A}(F(R))^{\frac12}
\end{eqnarray}
with $N_0$ the normalization constant. Concerning the localization of the scalar zero mode, in terms of
Eq. (\ref{ExprCondLocaZM}), there is
\begin{eqnarray} \label{IntZMSca}
  \int\chi_0^2dz=N_0^2\int e^{3A}F(R)dz=N_0^2\int e^{2A}F(R)dy.
\end{eqnarray}
Furthermore, the integrand above exhibits the following
asymptotic behaviors as $y\rightarrow+\infty$:
\begin{eqnarray} \label{AsymZMSca}
  e^{2A}F(R)\rightarrow 2^be^{t_1-bcy-(\frac54)^{t_2}t_1(\frac{b}{4+5b})^{t_2}e^{2t_2cy}}.
\end{eqnarray}
In the r.h.s of this expression, term $bcy$ diverges to positive infinity when $y\rightarrow+\infty$.
The asymptotic behavior of term $(\frac54)^{t_2}t_1(\frac{b}{4+5b})^{t_2}e^{2t_2cy}$ is decided by
parameter $t_2$: if $t_2<0$, this term tends to zero, and if $t_2>0$, it diverges to positive
infinity. Thus, without any constraints on parameter $t_2$, the above expression (\ref{AsymZMSca})
converges to zero when $y\rightarrow+\infty$, and the scalar zero mode can always be localized on
the thick brane.

In light of Eq. (\ref{OriZMMin}), the localized scalar zero mode is suppressed to zero at the brane position
when $t_2<0$. This condition brings about a novel perspective for the localization of scalar field, which
will be presented in the last of this section.

For the massive KK modes, we shall turn back to the Schr\"{o}dinger-like equation (\ref{SchroSca}). Since
there is no analytical expression for the effective potential (\ref{effVSca}) concerning coordinate $z$.
Based on the coordinate transformation (\ref{coordinate trans}), the effective potential can be expressed
in terms of coordinate $y$ as
\begin{eqnarray} \label{Potential ySca}
 V_0(z(y))=\frac{15}{4}{A'}^2e^{2A}+\frac{3}{2}{A''}e^{2A}
        +\frac{{F''}(R)e^{2A}}{2F(R)}+\frac{2A'{F'}(R)e^{2A}}{F(R)}
        -\frac{{F'}^2(R)e^{2A}}{4F^2(R)},
\end{eqnarray}
where the prime denotes the derivative with respect to $y$. By substituting the warp factor (\ref{Warped factor})
and the function $F(R)$ (\ref{FR5}) into the above expression, we can get
\begin{eqnarray} \label{EffPotenSca}
  V_0(z(y))&=&\frac{c^2}{16}(\text{sech}(cy))^{2+b}\bigg(\frac{(4+5b)^2\text{sech}^4(cy)}{b^2}\bigg)^{-t_2}
    \bigg[16\times25^{t_2}t_1^2t_2^2\sinh^2(cy)                           \nonumber    \\
   & &+3b\bigg(\frac{(4+5b)^2\text{sech}^4(cy)}{b^2}\bigg)^{t_2}(-4+5b\sinh^2(cy))       \nonumber     \\
   & &-16\times5^{t_2}t_1t_2\bigg(\frac{(4+5b)^2\text{sech}^4(cy)}{b^2}\bigg)^{t_2/2}(1+2(t_2-b)\sinh^2(cy))\bigg].\hspace{0.1cm}
\end{eqnarray}
From this expression, the behaviors of the effective potential at position $y=0$ and positive infinity are:
\begin{eqnarray}
  V_0(z(y=0)) &=& \frac14c^2\bigg[-3b-4t_1t_2\bigg(\frac{5b}{4+5b}\bigg)^{t_2}\bigg],  \label{V0Sca}    \\
  V_0(z(y\rightarrow+\infty)) &\rightarrow& \left\{
    \begin{array}{llcll}
      +\infty             & \hspace{0.8cm}  t_2 > b/4,     \\
      \mathrm{C}_0        & \hspace{0.8cm}  t_2 = b/4,     \\
      0                   & \hspace{0.8cm}  t_2 < b/4                        \label{AsymEffPotenSca}
\end{array} \right.
\end{eqnarray}
with the limit
\begin{eqnarray} \label{LmtSca}
  \mathrm{C}_0=\frac{1}{16}b^2c^2t_1^2\bigg(\frac{5b}{5b+4}\bigg)^{b/2}.
\end{eqnarray}
The former expression (\ref{V0Sca}) indicates that if $t_2\rightarrow0^-$, the effective potential is negative
at position $y=0$, and if $t_2\ll0$, it becomes positive.

Concerning the latter expression (\ref{AsymEffPotenSca}), the condition $t_2<b/4$ encompasses the case $t_2<0$,
which is permissible as the scalar zero mode can still be localized. Along with the zero mode, the effective
potential of this particular case will be investigated separately.

\begin{figure} 
\begin{center}
\subfigure[$V_{0}(z)$.]{\label{FigeffVSca}
\includegraphics[width= 0.48\textwidth]{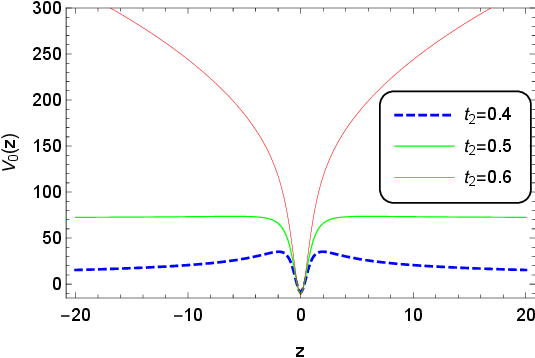}}
\subfigure[$\chi_0(z)$.]{\label{FigZMSca}
\includegraphics[width= 0.48\textwidth]{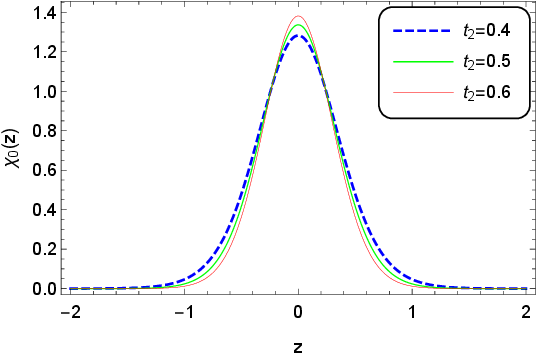}}
\end{center}\vskip -5mm
\caption{For $\mathcal{M}_4$ brane case, the effective potentials $V_{0}(z)$ in (a), and the shapes
         of scalar zero mode $\chi_0(z)$ in (b). The parameters are set as $b=2,c=1,t_1=20$, and
         $t_{2}=0.4,0.5,0.6$.}
 \label{FigZM-EPSca}
\end{figure}

When $t_2>0$, our focus is on the cases where the values of $t_2$ span the vicinity of the critical value $b/2$.
For example, the effective potential $V_0(z)$ and the scalar zero mode $\chi_0(z)$, depicted in Fig. \ref{FigZM-EPSca},
are obtained through numerical methods with parameters $b=2,c=1,t_1=20$, and various values of $t_2$: $0.4,0.5$,
and $0.6$. The figure demonstrates that the behaviors of the effective potential align well with the conclusions
(\ref{AsymEffPotenSca}). Subsequently, we will explicitly outline the localization of the massive KK modes in
the three cases of parameter $t_2$.

Initially, for the case of $t_2=0.4<b/4$, the effective potential adopts a volcanic profile,
enabling the quasi-localization of massive KK modes on the thick brane. These quasi-localized
modes, referred to as resonant KK modes, describe massive 4D scalars with finite lifetimes on
the brane \cite{QTanMPJC2383}. The resonant KK modes can be studied by the relative probability
method, which was proposed in Refs. \cite{YXLiuPRD0980,YXLiuPRD0980-2}. This method defines
relative probability as:
\begin{eqnarray} \label{DefReson}
  P_S(m^2)=\frac{\int^{z_b}_{-z_b}|\chi(z)|^2dz}{\int^{z_{max}}_{-z_{max}}|\chi(z)|^2dz},
\end{eqnarray}
where $2z_b$ is approximately the width of the thick brane, and $z_{max}=10z_b$. It is clear
that the KK modes tend to plane waves and the corresponding probability $P_S(m^2)$ approaches
$1/10$ when $m^2\gg V_{max}$ ($V_{max}$ is the maximum of the corresponding potential). The
lifetime $\tau$ of a resonant state is $\tau\sim\Gamma^{-1}$ with $\Gamma=\delta m$ being the
full width at half maximum of the resonant peak.

\begin{figure} 
\begin{center}
\subfigure[$t_1=25,t_2=0.3$.]{\label{FigResSpecSca}
\includegraphics[width= 0.35\textwidth]{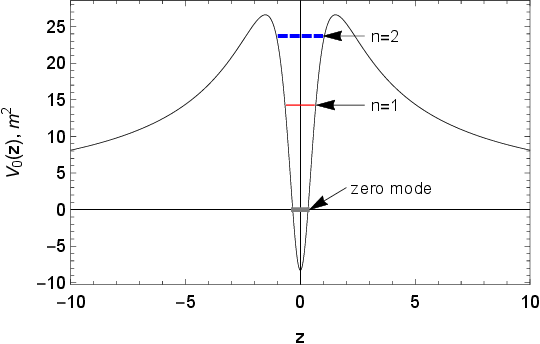}}
\hspace{0.5cm}
\subfigure[$t_1=25,t_2=0.3$.]{\label{FigResPSca}
\includegraphics[width= 0.35\textwidth]{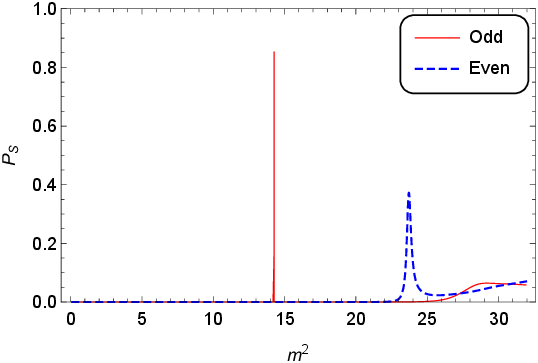}}
\subfigure[$t_1=20,t_2=0.4$.]{\label{2FigResSpecSca}
\includegraphics[width= 0.35\textwidth]{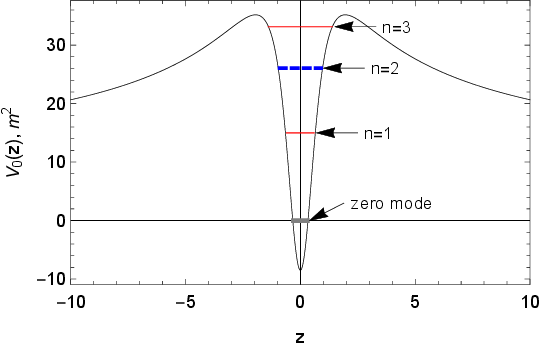}}
\hspace{0.5cm}
\subfigure[$t_1=20,t_2=0.4$.]{\label{2FigResPSca}
\includegraphics[width= 0.35\textwidth]{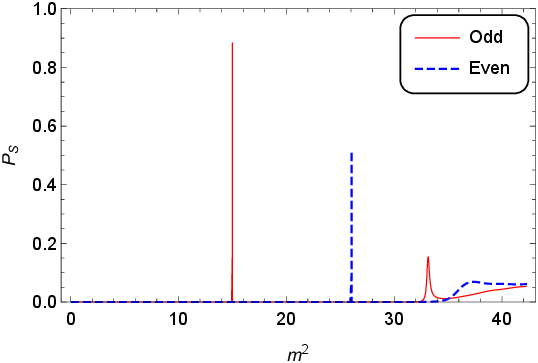}}
\subfigure[$t_1=25,t_2=0.4$.]{\label{3FigResSpecSca}
\includegraphics[width= 0.35\textwidth]{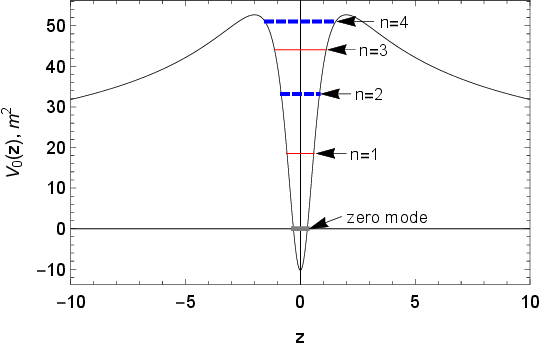}}
\hspace{0.5cm}
\subfigure[$t_1=25,t_2=0.4$.]{\label{3FigResPSca}
\includegraphics[width= 0.35\textwidth]{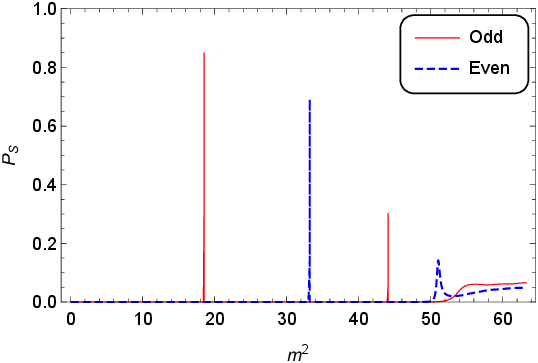}}
\end{center}\vskip -5mm
\caption{For $\mathcal{M}_4$ brane case, the mass spectra, the effective potential $V_0(z)$,
         and corresponding relative probability $P_S$ with parameter sets: $t_1=25,t_2=0.3$;
         $t_1=20,t_2=0.4$; and $t_1=25,t_2=0.4$. The potential $V_0(z)$ for the black
         line, the zero mode for the grey line, the even parity resonant KK modes for the blue
         lines, and the odd parity resonant KK modes for the red lines. The parameters are set
         as $b=2$ and $c=1$.}
 \label{FigResSpec-PSca}
\end{figure}

In this case where $t_2<b/4$, Fig. \ref{FigResSpec-PSca} illustrates the relative possibilities
corresponding to the parameters sets: $t_1=25,t_2=0.3$; $t_1=20,t_2=0.4$; and $t_1=25,t_2=0.4$.
Each peak in these figures represents a resonant KK mode. Additionally, the mass spectra alongside
the effective potentials are also shown in Fig. \ref{FigResSpec-PSca}. Detailed information regarding
the mass, width, and lifetime of all scalar resonant KK modes is provided in Table \ref{tableRMVolSca}.

\begin{figure} 
\begin{center}
\subfigure[$\chi_1$.]{\label{fig_Odd1Vol}
\includegraphics[width = 0.23\textwidth]{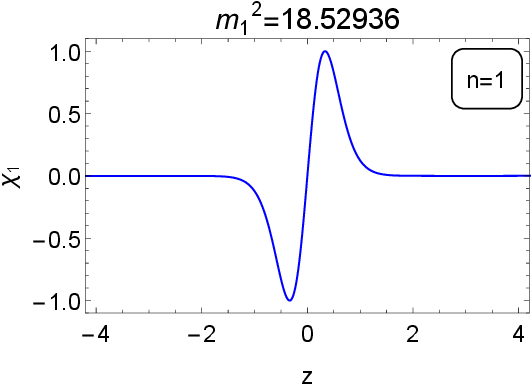}}
\subfigure[$\chi_2$.]{\label{fig_Even1Vol}
\includegraphics[width = 0.23\textwidth]{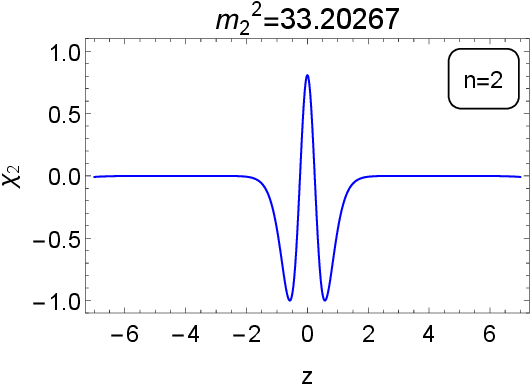}}
\subfigure[$\chi_3$.]{\label{fig_Odd2Vol}
\includegraphics[width = 0.23\textwidth]{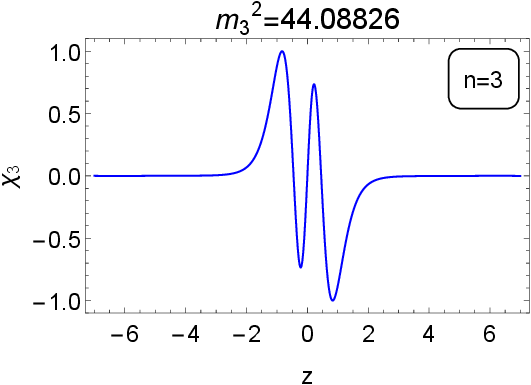}}
\subfigure[$\chi_4$.]{\label{fig_Even2Vol}
\includegraphics[width = 0.23\textwidth]{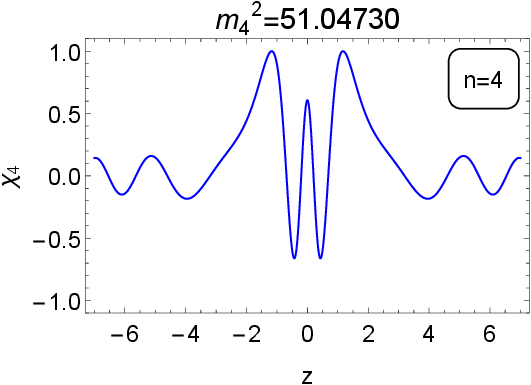}}
\end{center}\vskip -2mm
\caption{For $\mathcal{M}_4$ brane case, the shapes of resonance KK modes $\chi(z)$ for scalars
         with different $m^{2}$. The parameters are set as $b=2,c=1,
         t_1=25$ and $t_{2}=0.4$.}
 \label{FigResSca}
\end{figure}

Analysis of the mass spectra for the scalar KK modes (in Fig. \ref{FigResSpec-PSca}) reveals
that the ground state corresponds to the zero mode (a bound state), and all the massive KK modes
are resonant modes. Notably, the number of the resonant modes increases with both the parameters
$t_1$ and $t_2$. Further examination of Table \ref{tableRMVolSca} and Fig. \ref{3FigResPSca}
reveals the existence of four resonant KK modes when $t_1=25$ and $t_2=0.4$, visually depicted in
Fig. \ref{FigResSca}. Thus, when $t_2<b/4$, the scalar zero mode can be localized on the brane, and
the massive KK modes can be quasi-localized on the brane.

\begin{table}[tbp]
\centering
\begin{tabular}{|c|c|c|c|c|c|c|c|}
    \hline
    $t_1$               & $t_2$                 & $V^{\text{max}}_{0}$  & $n$         & $m^2$ &
    $m$                 & $\Gamma$              & $\tau$
    \\
    \hline
    $25$                & $0.3$                 & $26.6025$             & $1$         & $14.2663$ &
    $3.7771$            & $1.247\times10^{-5}$  & $8.022\times10^4$
    \\
                        &                       &                       & $2$         & $23.7144$ &
    $4.8698$            & $0.0382$              & $26.1855$
    \\
    \hline
    $20$                & $0.4$                 & $35.1828$             & $1$         & $14.9973$ &
    $3.8726$            & $1.488\times10^{-10}$ & $6.718\times10^9$
    \\
                        &                       &                       & $2$         & $26.0604$ &
    $5.1049$            & $1.023\times10^{-5}$  & $9.775\times10^4$
    \\
                        &                       &                       & $3$         & $33.1468$ &
    $5.75733$           & $0.0261$              & $38.3825$
    \\
    \hline
    $25$                & $0.4$                 & $52.6928$             & $1$         & $18.5294$ &
    $4.3046$            & $1.000\times10^{-14}$ & $1.024\times10^{14}$
    \\
                        &                       &                       & $2$         & $33.2027$ &
    $5.7677$            & $1.699\times10^{-9}$  & $5.885\times10^8$
    \\
                        &                       &                       & $3$         & $44.0883$ &
    $6.6399$            & $1.541\times10^{-4}$  & $6.488\times10^3$
    \\
                        &                       &                       & $4$         & $51.0473$ &
    $7.14474$           & $0.0402$              & $24.8919$
    \\
    \hline
\end{tabular}
\caption{For $\mathcal{M}_4$ brane case, the mass, width, and lifetime of resonant KK modes of the
        scalar fields. The parameters are set as $b=2$ and $c=1$. }
    \label{tableRMVolSca}
\end{table}

\begin{figure} 
\begin{center}
\includegraphics[width= 0.49\textwidth]{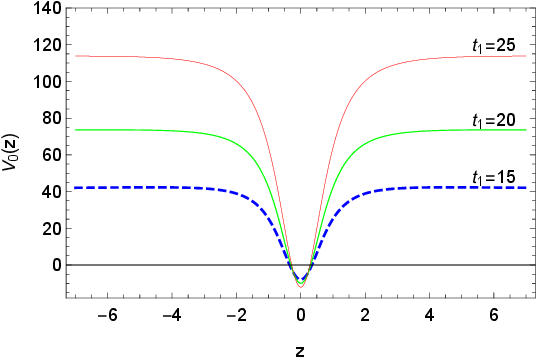}
\caption{For $\mathcal{M}_4$ brane case, the effective potential $V_0(z)$ with the parameter $t_1=15,20$, and $25$.
        The other parameters are set as $b=2,c=1$, and $t_2=0.5$.}
\label{figSpecPTSca}
\end{center}
\end{figure}

In the case where $t_2$ equals $0.5$, equivalent to $b/4$, a P\"{o}schl-Teller potential emerges,
enabling the localization of a finite number of massive KK modes on the thick brane. The limit
$\text{C}_0$ (\ref{LmtSca}) of the effective potential is dependent on parameters $b,c$ and $t_1$.
Our focus will center on the influence of parameter $t_1$. The shapes of the effective potentials
are shown in Fig. \ref{figSpecPTSca}, with the parameter $t_1$ set at values $15,20$, and $25$.
Notably, the figure illustrates a series of potential wells, which become deeper
with increasing $t_1$, leading to a greater number of localized massive KK modes. Specifically,
the mass spectra of the localized KK modes are listed as follows:
\begin{eqnarray}
  m_n^2=&\hspace{-3.50cm}\{0,13.96,24.63,32.37,37.58\},                     &\text{for}~t_1=15,    \label{SpecPTSca1} \\
  m_n^2=&\hspace{-0.38cm}\{0,18.23,33.24,45.30,54.70,61.77,66.85,70.29\},   &\text{for}~t_1=20,    \label{SpecPTSca2} \\
  m_n^2=&\{0,22.48,41.79,58.13,71.74,82.86,91.77,98.73,~\ ~\                &\nonumber         \\
        &\hspace{-4.2cm}104.02,107.90,110.63\},                               &\text{for}~t_1=25.     \label{SpecPTSca3}
\end{eqnarray}
From these mass spectra, it is apparent that when $t_2$ equals $b/4$, the zero mode is localized
on the thick brane, and the number of the localized massive KK modes increases with the parameter
$t_1$.

\begin{figure} 
\begin{center}
\includegraphics[width= 0.49\textwidth]{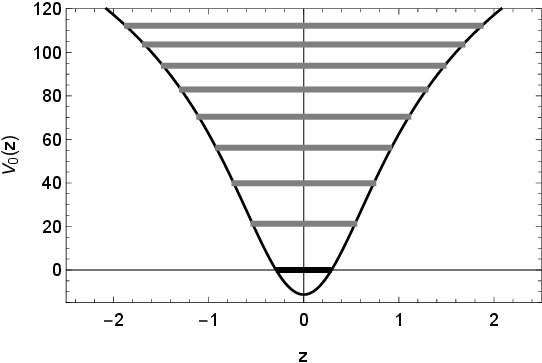}
\caption{For $\mathcal{M}_4$ brane case, the shape of the effective potential $V_0(z)$; the thick black line
        corresponds to $m_0^2=0$ and the first eight massive levels ($1\leq n\leq8$)
        of the $m_n^2$ spectrum are given by the grey lines. The parameter are
        set as $b=2,c=1,t_1=20$ and $t_2=0.6$.}
\label{figSpecIDWSca}
\end{center}
\end{figure}

Lastly, for the case of $t_2=0.6>b/4$, there is an infinitely deep well, which confines all massive
KK modes on the thick brane. These localized modes form an infinite discrete spectrum of mass.
The effective potential and the mass spectrum of lower localized KK modes are shown in Fig. \ref{figSpecIDWSca}
with specific values of parameters.

Therefore, when $t_2>0$, the scalar zero mode is localized on the thick brane, and the massive KK modes exhibit
localization or quasi-localization on the brane.

Then, for the case of $t_2<0$, it has been determined that scalar zero mode is localized, and from Eq. (\ref{V0Sca}),
the effective potential can exhibit positivity at the origin of the extra dimension. This positivity could give
rise to the presence of a local minimum of the scalar zero mode. At the origin of the extra dimension, the scalar
zero mode (\ref{zeromodeSca}), as well as its first-order and second-order derivatives, are
\begin{eqnarray}
  \chi_0(z(y=0)) &=& \sqrt{e^{t_1[1-(\frac{5b}{5b+4})^{t_2}]}},                                        \label{OriZMSca}          \\
  \partial_y\chi_0(z(y=0)) &=& 0,                                                                      \label{1stDeriOriZMSca}   \\
  \partial_y^2\chi_0(z(y=0)) &=& -\frac{c^2}{4}\sqrt{e^{t_1[1-(\frac{5b}{5b+4})^{t_2}]}}
        \bigg[3b+4t_1t_2\bigg(\frac{5b}{5b+4}\bigg)^{t_2}\bigg].                                       \label{2stDeriOriZMSca}
\end{eqnarray}
From Eq. (\ref{OriZMSca}), we can infer that if $t_2\ll0$, the zero mode will be suppressed to zero
at the origin of the extra dimension. According to Eq. (\ref{1stDeriOriZMSca}), the zero mode could
have a local extremum at position $z=0$. Further verification of this extremum depends on the behavior
described by the second derivative (\ref{2stDeriOriZMSca}).

By setting $\partial_y^2\chi_0(z(y=0))=0$, we can obtain the following critical value $t_{2\text C}$
for the parameter $t_2$:
\begin{eqnarray}  \label{t22ndDeriOriZMSca}
  t_{2\text C}=\frac{\text{ProductLog}[\text C_1,-\frac{3b}{4t_1}\ln[\frac{5b}{5b+4}]]}{\ln[\frac{5b}{5b+4}]}
\end{eqnarray}
where ProductLog is the Lambert $W$ function, and C$_1$ is an integer. Then, at the origin of the extra
dimension, if $t_{2\text C}<t_2<0$, Eq. (\ref{2stDeriOriZMSca}) turns out to be negative, indicating that
the zero mode has a local maximum. If $t_2=t_{2\text C}$, Eq. (\ref{2stDeriOriZMSca}) equals zero, so the
zero mode exhibits neither a maximum nor a minimum. Lastly, if $t_2<t_{2\text C}$, Eq. (\ref{2stDeriOriZMSca})
becomes positive, resulting in a local minimum of the zero mode.

As the Schr\"{o}dinger-like equation (\ref{Sch-equa}) is given in terms of coordinate $z$. Based on the
coordinate transformation (\ref{coordinate trans}), we can get
\begin{eqnarray}
  \partial_z\chi   &=& \partial_y\chi\partial_zy,                                        \label{1stZMCoorTrans}    \\
  \partial_z^2\chi &=& \partial_y^2\chi(\partial_zy)^2+\partial_y\chi\partial_z^2y.      \label{2ndZMCoorTrans}
\end{eqnarray}
So, there is
\begin{eqnarray}
  \partial_y\chi(z(y=0))=0 &\Rightarrow& \partial_z\chi(z=0)=0,                   \label{ZP1stZMDeriZMSca}   \\
  \left.
  \begin{matrix}
    \partial_y\chi_0(z(y=0))=0     \vspace{0.1cm}                     \\
    \partial_y^2\chi_0(z(y=0))=0
  \end{matrix}
  \right\}
     &\Rightarrow& \partial_z^2\chi_0(z=0)=0.                                     \label{ZP2nd2ndDeriZMSca}
\end{eqnarray}
At the origin of the extra dimension, the type of extremum of the zero mode remains unchanged regardless
of there is coordinate $y$ or $z$.

\begin{figure} 
\begin{center}
\subfigure[$V_0(z)$.]{\label{FigCritt2EffPotOriSca}
\includegraphics[width= 0.45\textwidth]{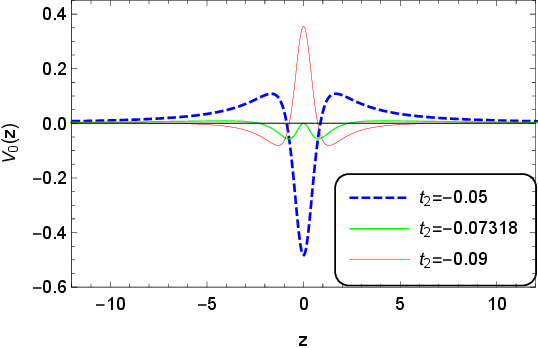}}
\hspace{0.1cm}
\subfigure[$\chi_0(z)$.]{\label{FigCritt2ZMOriSca}
\includegraphics[width= 0.45\textwidth]{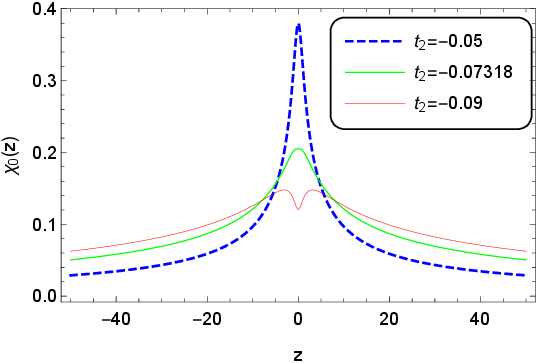}}
\end{center}\vskip -5mm
\caption{For $\mathcal{M}_4$ brane case, the effective potential $V_0(z)$ in (a) and zero mode $\chi_0(z)$ in (b).
         The parameters are set as $b=2,c=1$, and $t_{1}=20$.}
 \label{FigCritt2ZMEffPotSca}
\end{figure}

On the other hand, the Schr\"{o}dinger-like equation (\ref{Sch-equa}) can be rewritten as
\begin{eqnarray} \label{TransShroEqtSca-1}
   \partial_z^2\chi_n(z)=[V_0(z)-m_n^2]\chi_n(z).
\end{eqnarray}
For the zero mode, $m_0=0$, this expression becomes
\begin{eqnarray} \label{ZMTransShroEqtSca-1}
   \partial_z^2\chi_0(z)=V_0(z)\chi_0(z).
\end{eqnarray}
So, if $\partial_z^2\chi_0(z=0)$, there is $V_0(z=0)=0$. This can
also be observed from Eqs. (\ref{V0Sca}) and (\ref{2stDeriOriZMSca}), as the former acts as a factor of the
latter. This relation suggests that at the origin of the extra dimension, if the negative effective potential
becomes positive, the local maximum of the zero mode turns into the local minimum. In Fig. \ref{FigCritt2ZMEffPotSca},
the effective potential and the zero mode are plotted via numerical methods with certain values of
parameters. It can be seen that as the parameter $t_2$ decreases, and becomes less than the critical value
$t_{2\text C}$, a potential barrier will emerge at the origin of the extra dimension, giving rise to a local
minimum of the zero mode.

\begin{figure} 
\begin{center}
\subfigure[$V_0(z)$.]{\label{FigEffPotOriSca}
\includegraphics[width= 0.45\textwidth]{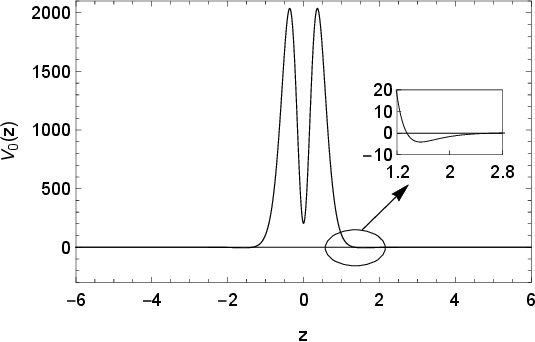}}
\hspace{0.1cm}
\subfigure[$\chi_0(z)$.]{\label{FigZMOriSca}
\includegraphics[width= 0.45\textwidth]{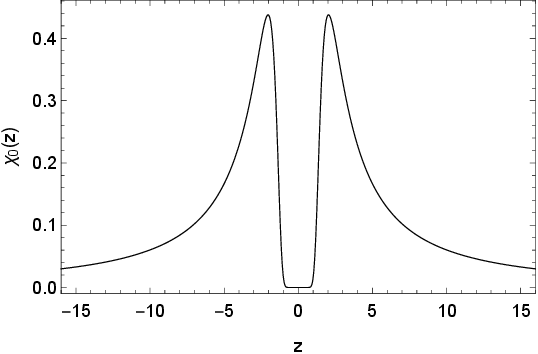}}
\end{center}\vskip -5mm
\caption{For $\mathcal{M}_4$ brane case, the effective potential $V_0(z)$ in (a) and zero mode $\chi_0(z)$
         for the scalar field in (b). The parameters are set as $b=2,c=1,t_1=25$, and $t_{2}=-3$.}
 \label{FigZMEffPotSca}
\end{figure}

Additionally, for Eq. (\ref{OriZMSca}), the zero mode tends to $0^+$ as $t_2\ll0$ at position $y=0$.
In Fig. \ref{FigZMEffPotSca}, both the effective potential and the zero mode are depicted, with lower
value of parameter $t_2$. As shown in Fig. \ref{FigZMOriSca}, the scalar zero mode is suppressed to
zero at position $z=0$, and appears localized on both sides of the origin of the extra dimension.

For the effective potential, Fig. \ref{FigEffPotOriSca} illustrates the presence of a positive potential
well with a positive lower boundary at the brane position. This well is flanked by two comparatively
shallower negative potential wells. The existence of negative potential wells facilitates the localization
of the scalar zero mode, while the positive potential well potentially contributes to the quasi-localization
of the massive KK modes.

\begin{figure} 
\begin{center}
\subfigure[$t_1=25,t_2=-2$.]{\label{FigResSpecOriSca}
\includegraphics[width= 0.35\textwidth]{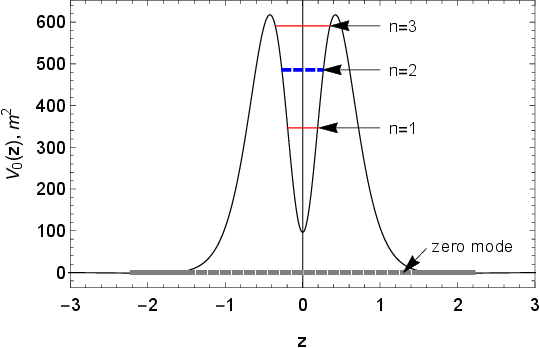}}
\hspace{0.5cm}
\subfigure[$t_1=25,t_2=-2$.]{\label{FigResPOriSca}
\includegraphics[width= 0.35\textwidth]{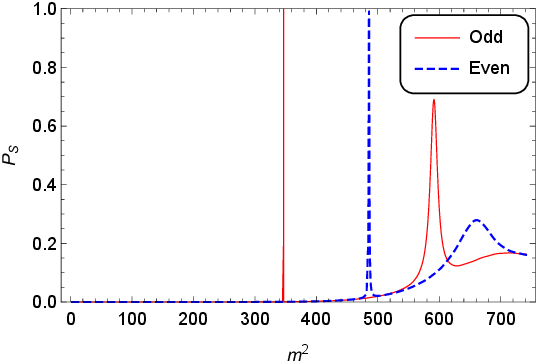}}
\subfigure[$t_1=20,t_2=-3$.]{\label{2FigResSpecOriSca}
\includegraphics[width= 0.35\textwidth]{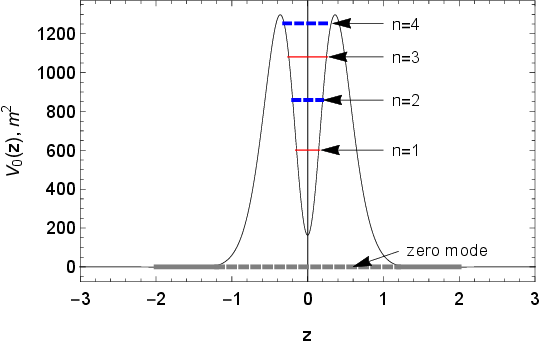}}
\hspace{0.5cm}
\subfigure[$t_1=20,t_2=-3$.]{\label{2FigResPOriSca}
\includegraphics[width= 0.35\textwidth]{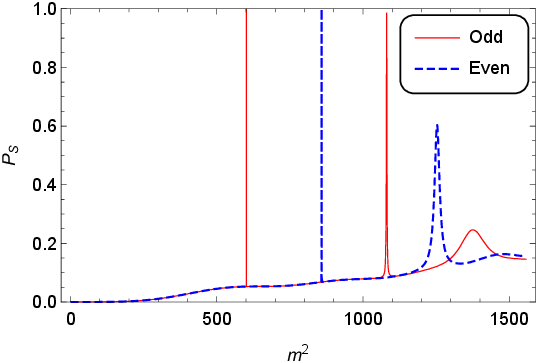}}
\subfigure[$t_1=25,t_2=-3$.]{\label{3FigResSpecOriSca}
\includegraphics[width= 0.35\textwidth]{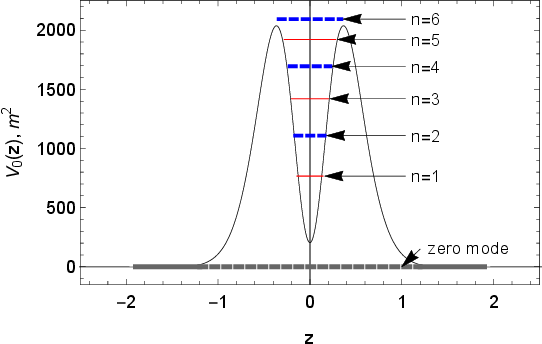}}
\hspace{0.5cm}
\subfigure[$t_1=25,t_2=-3$.]{\label{3FigResPOriSca}
\includegraphics[width= 0.35\textwidth]{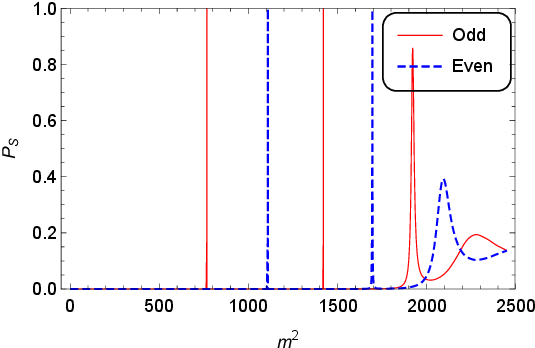}}
\end{center}\vskip -5mm
\caption{For $\mathcal{M}_4$ brane case, the mass spectra, the effective potential $V_0(z)$, and corresponding
         relative probability $P_S$ with parameter sets: $t_1=25,t_2=-2$; $t_1=20,t_2=-3$;
         and $t_1=25,t_2=-3$. $V_0(z)$ for the black line, the zero mode for the grey line
         (including the grey dashed line), the even parity resonant KK modes for the blue lines, and
         the odd parity resonant KK modes for the red lines. The parameters are set as $b=2$ and $c=1$.}
 \label{FigResSpec-POriSca}
\end{figure}

Concerning the positive potential well in Fig. \ref{FigEffPotOriSca}, the solutions for the
resonant KK modes can be obtained using numerical methods. Figure \ref{FigResSpec-POriSca}
showcases the profiles of the relative probability $P_S$ for various values of parameters $t_1$ and
$t_2$. The corresponding mass spectra alongside the effective
potentials are also displayed in Fig. \ref{FigResSpec-POriSca}, while Table \ref{tableRMOriSca} lists
the mass, width, and lifetime of all scalar resonant KK modes.

\begin{figure} 
\begin{center}
\subfigure[$\chi_1$.]{\label{fig_Odd1Ori}
\includegraphics[width = 0.30\textwidth]{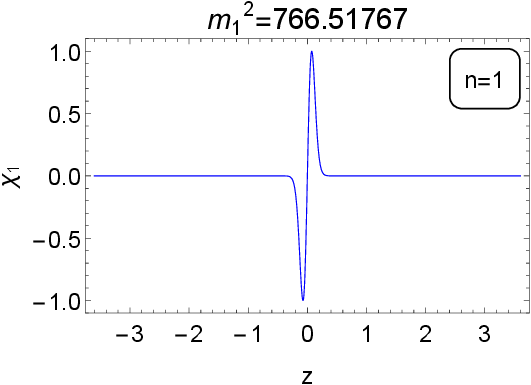}}
\subfigure[$\chi_2$.]{\label{fig_Even1Ori}
\includegraphics[width = 0.30\textwidth]{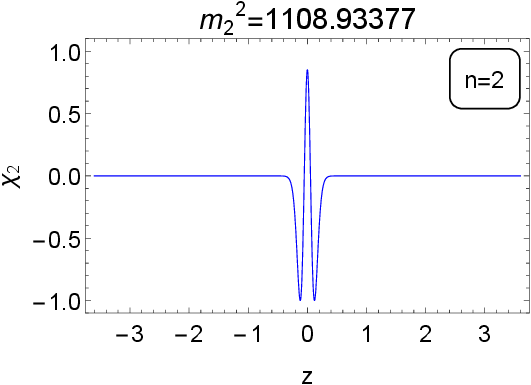}}
\subfigure[$\chi_3$.]{\label{fig_Odd2Ori}
\includegraphics[width = 0.30\textwidth]{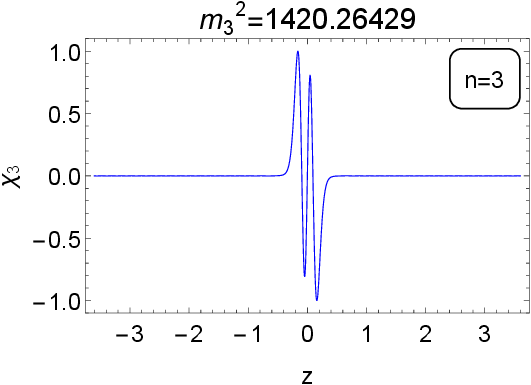}}
\subfigure[$\chi_4$.]{\label{fig_Even2Ori}
\includegraphics[width = 0.30\textwidth]{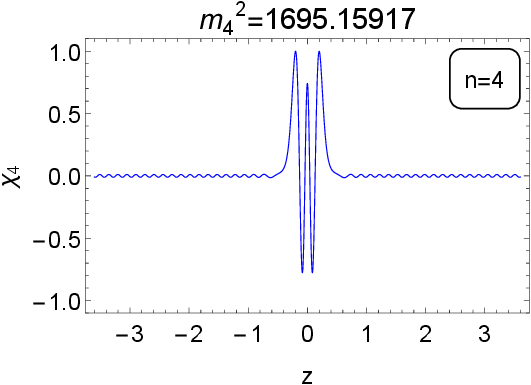}}
\subfigure[$\chi_5$.]{\label{fig_Odd3Ori}
\includegraphics[width = 0.30\textwidth]{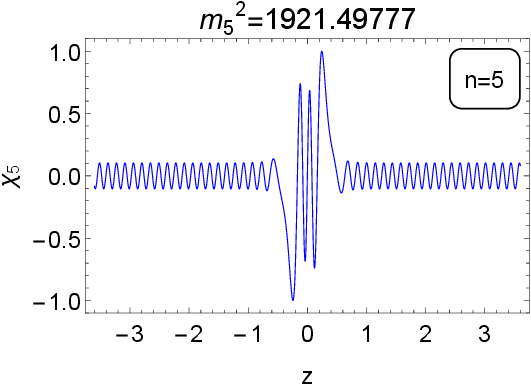}}
\subfigure[$\chi_6$.]{\label{fig_Even3Ori}
\includegraphics[width = 0.30\textwidth]{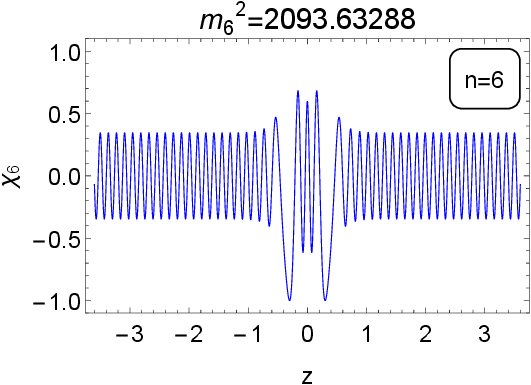}}
\end{center}\vskip -2mm
\caption{For $\mathcal{M}_4$ brane case, the shapes of resonant KK modes for scalars with different
         $m^{2}$. The parameters are set as $b=2,c=1,t_1=25$ and
         $t_{2}=-3$.}
 \label{FigResOriSca}
\end{figure}

Observations from Fig. \ref{FigResSpec-POriSca} and Table \ref{tableRMOriSca} indicate an increase in
the number of the resonant KK modes with higher values of $t_1$, or decreasing values of $t_2$. Besides,
in Fig. \ref{FigResSpec-POriSca}, the effective potential displays two negative potential wells surrounding
the origin of the extra dimension, resulting in the localization of the zero mode on both sides of the
origin. In contrast, the positive potential well situated at the brane position leads to the quasi-localization
of the massive KK modes. Figure \ref{FigResOriSca} displays all resonant KK modes for the case of $t_1=25$
and $t_2=-3$, indicating the quasi-localization of these massive modes.

\begin{table}[tbp]
\centering
\begin{tabular}{|c|c|c|c|c|c|c|c|}
    \hline
    $t_1$               & $t_2$                 & $V^{\text{max}}_{0}$  & $n$         & $m^2$ &
    $m$                 & $\Gamma$              & $\tau$
    \\
    \hline
    $25$                & $-2$                  & $617.559$             & $1$         & $346.5786$ &
    $18.6166$           & $1.017\times10^{-3}$  & $983.3708$
    \\
                        &                       &                       & $2$         & $485.6267$ &
    $22.0369$           & $0.0292$              & $34.2018$
    \\
                        &                       &                       & $3$         & $591.2284$ &
    $24.3151$           & $0.3248$              & $3.0787$
    \\
    \hline
    $20$                & $-3$                  & $1.298\times10^3$     & $1$         & $601.0303$ &
    $24.5159$           & $3.551\times10^{-5}$  & $2.816\times10^4$
    \\
                        &                       &                       & $2$         & $858.7666$ &
    $29.3047$           & $1.663\times10^{-3}$  & $601.3832$
    \\
                        &                       &                       & $3$         & $1.081\times10^3$ &
    $32.8780$           & $0.0316$              & $31.6267$
    \\
                        &                       &                       & $4$         & $1.253\times10^3$ &
    $35.4042$           & $0.3567$              & $2.8038$
    \\
    \hline
    $25$                & $-3$                  & $2.037\times10^3$     & $1$         & $766.5177$ &
    $27.6861$           & $4.407\times10^{-7}$  & $2.269\times10^6$
    \\
                        &                       &                       & $2$         & $1.109\times10^3$ &
    $33.3007$           & $3.661\times10^{-5}$  & $2.732\times10^4$
    \\
                        &                       &                       & $3$         & $1.420\times10^3$ &
    $37.6864$           & $1.303\times10^{-3}$  & $767.2625$
    \\
                        &                       &                       & $4$         & $1.695\times10^3$ &
    $41.1723$           & $0.0232$              & $43.1549$
    \\
                        &                       &                       & $5$         & $1.921\times10^3$ &
    $43.8349$           & $0.2083$              & $4.7992$
    \\
                        &                       &                       & $6$         & $2.094\times10^3$ &
    $45.7562$           & $1.1316$              & $0.8837$
    \\
    \hline
\end{tabular}
\caption{For $\mathcal{M}_4$ brane case, the mass, width, and lifetime of resonant KK modes of
        the scalar fields. The parameters are set as $b=2$ and $c=1$. }
    \label{tableRMOriSca}
\end{table}

\begin{figure} 
\begin{center}
\includegraphics[width= 0.96\textwidth]{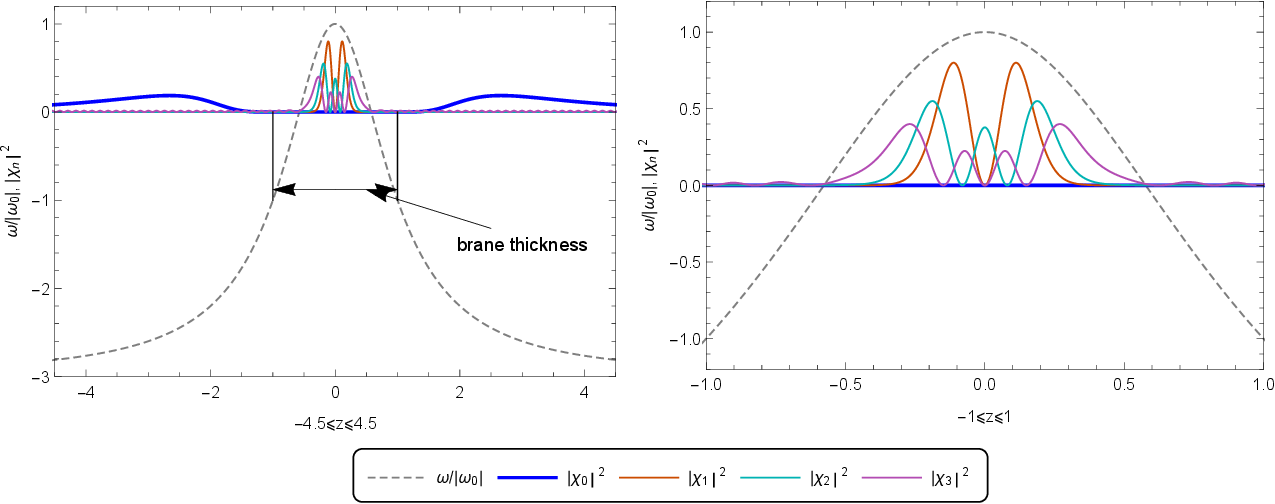}
\caption{For $\mathcal{M}_4$ brane case, the relative energy density $\omega/|\omega_0|$ of the 5D
        RSII-like model for the dashed line, the probability densities $|\chi_n|^2$ of the scalar
        zero mode and the scalar resonant modes for the colored lines. The parameters are set as
        $b=2,c=1,t_1=25$ and $t_2=-2$.}
\label{figZMResOriSca}
\end{center}
\end{figure}

Upon examining the profiles of the zero mode and the resonant KK modes, distinct probability density
distributions become apparent along the extra dimension, as illustrated in Fig. \ref{figZMResOriSca}. Besides,
the energy density $\omega$ of the thick brane can be given by
\begin{eqnarray} \label{DensEnerg}
   \omega=-g^{00}T_{00}
\end{eqnarray}
with $T_{00}$ the component of the energy-momentum tensor of brane world. The relative energy density
$\omega/|\omega_0|$ with $\omega_0=\omega(z=0)$ for the 5D RSII-like model (\ref{Warped factor}) is
also depicted in the same figure, identifying the position of the brane. As usual, the brane thickness
is defined as the full width at half maximum of the peak of the energy density, and we also show it
in Fig. \ref{figZMResOriSca}. It can be seen that the scalar zero mode is localized on both sides of
the thick $\mathcal{M}_4$ brane, while the massive modes are quasi-localized at the origin of the
extra dimension. The splitting of the zero mode emerges, and the distribution of the zero mode and
resonant KK modes exhibits a separation along the extra dimension. Previous works \cite{Arkani-HamedPRD2000,
Arkani-HamedPRD2000115004,DCCaiPRD2006} have presented the so-called "split fermion" model, where
different fermion KK modes could be localized at different positions. Here, the scalar zero mode
and the massive modes can also be localized, or quasi-localized at different positions of the extra
dimension.

Therefore, in the scenario of $t_2<0$, the zero mode can be localized, and if parameter $t_2$ is less than
the critical value $t_{2\text C}$, the zero mode will be localized on both sides of the thick brane. Besides,
the massive modes can be quasi-localized on the origin of the extra dimension.

\subsubsection{$U(1)$ gauge vector fields} \label{flatVector}

The $U(1)$ gauge vector fields correspond to the $1-$form fields, and we denote a free bulk vector field
with $A_{\mu}$. This field is used to describe the particles with spin$-1$. In view of the gauge freedom,
we choose the gauge condition $A_z=0$ when conducting the KK decomposition.

Considering the coupling between the kinetic term and background spacetime, we assume the 5D
action for a free bulk vector field as
\begin{eqnarray} \label{ActVec}
 S_1=-\frac14\int d^5x\sqrt{-g}F(R)F_{MN}F^{MN},
\end{eqnarray}
where $F_{MN}=\partial_MA_N-\partial_NA_M$ is the field strength. Based on the flat metric
(\ref{metricMin}), the equations of motion corresponding to the aforementioned action can be derived
as follows:
\begin{eqnarray}
 \partial_{\mu}(\sqrt{-g}F(R)F^{\mu\nu})+\partial_z(\sqrt{-g}F(R)F^{z\nu})&=& 0,     \label{EoMVec1}    \\
    \partial_{\mu}(\sqrt{-g}F(R)F^{\mu z})&=& 0.      \label{EoMVec2}
\end{eqnarray}
By choosing the KK decomposition
\begin{eqnarray} \label{DecomVec}
 A_{\mu}(x^{\mu},z)=\sum_na^{(n)}_{\mu}(x^{\mu})\rho_n(z)e^{-\frac12A}(F(R))^{-\frac12},
\end{eqnarray}
the Schr\"{o}dinger-like equation for the KK modes $\rho_n(z)$ can be obtained as
\begin{eqnarray} \label{SchroVec}
 (-\partial^2_z+V_1(z))\rho_n(z)=m_n^2\rho_n(z),
\end{eqnarray}
where the effective potential $V_1(z)$ is
\begin{eqnarray} \label{EffPotVec}
 V_1(z)=\frac12A''+\frac14A'^2+\frac12\frac{A'F'(R)}{F(R)}+\frac{F''(R)}{2F(R)}-\frac{F'^2(R)}{4F^2(R)}
\end{eqnarray}
with $m_n$ the masses of the vector KK modes.

With the orthonormality conditions
\begin{eqnarray} \label{OrthoVec4}
  \int\rho_m(z)\rho_n(z)dz=\delta_{mn},
\end{eqnarray}
the full 5D action (\ref{ActVec}) can be reduced to the following 4D effective one
when integrated over the extra dimension:
\begin{eqnarray} \label{EffActVec4}
  S_1=-\frac14\sum_n\int d^4x\sqrt{-\hat g}(f_{\mu\nu}f^{\mu\nu}+2m_n^2a_{\mu}a^{\mu}),
\end{eqnarray}
where $f_{\mu\nu}=\partial_{\mu}a_{\nu}-\partial_{\nu}a_{\mu}$ is the 4D field strength.

From the Eq. (\ref{SchroVec}), the vector zero mode with $m_0^2=0$ can be solved as
\begin{eqnarray} \label{ZMVec}
  \rho_0(z)=N_1e^{\frac12A}(F(R))^{\frac12},
\end{eqnarray}
where $N_1$ is the normalization constant. Based on Eq. (\ref{ExprCondLocaZM}), the localization
condition for the vector zero mode can be expressed as
\begin{eqnarray} \label{LocZMVec}
  \int \rho_0^2(z)dz&=& N_1^2\int e^AF(R)dz       \nonumber     \\
    &=& N_1^2 \int F(R)dy=1.
\end{eqnarray}

In the minimal coupling case where $F(R)\equiv1$, it is evident that the above normalization condition
(\ref{LocZMVec}) cannot be met, and the vector zero mode is unnormalizable. Besides, with the
RSII-like model (\ref{Warped factor}) and the coordinate transformation (\ref{coordinate trans}),
the effective potential $V_1(z)$ (\ref{EffPotVec}) can be formulated in terms of coordinate $y$:
\begin{eqnarray} \label{EffPotVecMC}
 V_1(z(y))&=& \frac34A'^2e^{2A}+\frac12A''e^{2A}            \nonumber    \\
   &=& \frac{1}{16}bc^2(\text{sech}(cy))^{b+2}[3b(\sinh(cy))^2-4].
\end{eqnarray}
It is noticeable that this potential tends towards zero as $y\rightarrow+\infty$.
Consequently, neither the zero mode nor the massive KK modes can be localized on the thick brane.
To address this limitation, introducing coupling becomes necessary.

With consideration of the coupling function (\ref{FR5}), the asymptotic behavior of the integrand in the
normalization condition (\ref{LocZMVec}) as $y\rightarrow +\infty$ is
\begin{eqnarray} \label{AsymLocZMVec}
  F(R(y\rightarrow+\infty))\rightarrow e^{t_1-2^{-2t_2}t_1(\frac{5b}{4+5b})^{t_2}e^{2t_2cy}}.
\end{eqnarray}
It can be seen that when the parameter $t_2>0$, this integrand converges to zero double-exponentially as
$y\rightarrow +\infty$. Conversely, for $t_2<0$, it tends to $e^{t_1}$. Therefore, when $t_2>0$, the
normalization condition (\ref{LocZMVec}) is fulfilled, and the vector zero mode can be localized on the
thick brane.

For the massive vector KK modes, based on the coordinate transformation (\ref{coordinate trans}),
we substitute the warp factor (\ref{Warped factor}) and the function $F(R)$ (\ref{FR5}) into
the effective potential (\ref{EffPotVec}), and obtain
\begin{eqnarray} \label{SovEffPotVec}
  V_1(z(y))&=&\frac{c^2}{16}(\text{sech}(cy))^{2+b}\bigg(\frac{(4+5b)^2\text{sech}^4(cy)}{b^2}\bigg)^{-t_2}
    \bigg[16\times25^{t_2}t_1^2t_2^2\sinh^2(cy)                           \nonumber    \\
   & &+b\bigg(\frac{(4+5b)^2\text{sech}^4(cy)}{b^2}\bigg)^{t_2}(-4+3b\sinh^2(cy))       \nonumber     \\
   & &-16\times5^{t_2}t_1t_2\bigg(\frac{(4+5b)^2\text{sech}^4(cy)}{b^2}\bigg)^{t_2/2}(1-(b-2t_2)\sinh^2(cy))\bigg].
\end{eqnarray}
The behaviors of this effective potential at $y=0$ and positive infinity are:
\begin{eqnarray}
  V_1(z(y=0)) &=& \frac14c^2\bigg(-b-4t_1t_2\bigg(\frac{5b}{4+5b}\bigg)^{t_2}\bigg),  \label{y0EffPotVec}  \\
  V_1(z(y\rightarrow+\infty))&\rightarrow& \left\{
    \begin{array}{llcll}
      +\infty             & \hspace{0.8cm} t_2  > b/4,        \\
      \mathrm{C}_1        & \hspace{0.8cm} t_2  = b/4,        \\
      0                   & \hspace{0.8cm} 0<t_2  < b/4            \label{AsymEffPotenVec}
\end{array} \right.
\end{eqnarray}
with the limit
\begin{eqnarray} \label{LmtVec}
  \mathrm{C}_1=\frac{1}{16}b^2c^2t_1^2\bigg(\frac{5b}{5b+4}\bigg)^{b/2}.
\end{eqnarray}
For expression (\ref{y0EffPotVec}), it can be seen that the condition $V_1(z(y=0))>0$
necessitates a negative parameter $t_2$. However, this solution contradicts the requirement of $t_2>0$ for
localizing the vector zero mode. Consequently, $t_2$ must remain positive, and the effective potential
is consistently negative at the origin of the extra dimension.

\begin{figure} 
\begin{center}
\subfigure[$V_{1}(z)$.]{\label{FigeffVVec}
\includegraphics[width= 0.48\textwidth]{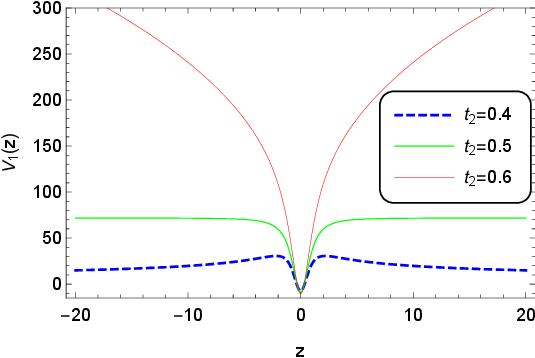}}
\subfigure[$\rho_0(z)$.]{\label{FigZMVec}
\includegraphics[width= 0.48\textwidth]{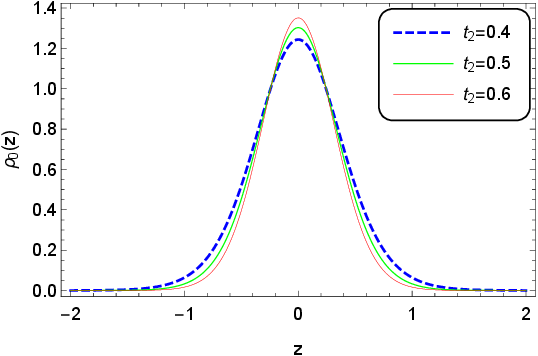}}
\end{center}\vskip -5mm
\caption{For $\mathcal{M}_4$ brane case, the effective potentials $V_{1}(z)$ in (a), and the shapes of the vector zero mode
         $\rho_0(z)$ in (b). The parameters are set as $b=2,c=1,t_1=20$, and $t_{2}=0.4,0.5,0.6$,
         respectively.}
 \label{FigZM-EPVec}
\end{figure}

The latter expression (\ref{AsymEffPotenVec}) illustrates that the asymptotic behaviors of potential
$V(z(y\rightarrow+\infty))$ closely resemble those observed for the scalar fields. The distinction is that
localizing the vector zero mode necessitates a positive parameter $t_2$, whereas localizing the scalar zero
mode does not. The effective potentials $V_1(z)$ and vector zero mode $\rho_0(z)$ are visually presented
in Fig. \ref{FigZM-EPVec} with different values of parameters. It is worth mentioning that the behaviors
of the vector zero mode and effective potential accord with the investigation in Ref \cite{ZHZhaoJHEP2018}.

Initially, when $t_2=0.4<b/4$, the effective potential takes on a volcano-shaped form.
No localized massive KK modes are present, but the resonant KK modes might exist on the
thick brane. Referring to the method outlined in Refs. \cite{YXLiuPRD0980,YXLiuPRD0980-2},
the relative probability function for a vector resonance on the thick brane is defined as
\begin{eqnarray} \label{RPResVec}
   P_V(m^2)=\frac{\int^{z_b}_{-z_b}|\rho(z)|^2dz}{\int^{z_{max}}_{-z_{max}}|\rho(z)|^2dz},
\end{eqnarray}
where $2z_b$ approximately represents the width of the thick brane, and $z_{max}$ is set as $10z_b$. For
the KK modes with significantly larger $m^2$ values than the maximum of the corresponding potential, they
tend to plane waves, resulting in the probabilities for them approaching $0.1$. The lifetime $\tau$ of
a resonant mode is $\tau\sim\Gamma^{-1}$, where $\Gamma=\delta m$ is the full width at half
maximum of the resonant peak.

\begin{figure} 
\begin{center}
\subfigure[$t_1=25,t_2=0.3$.]{\label{FigResSpecVec}
\includegraphics[width= 0.35\textwidth]{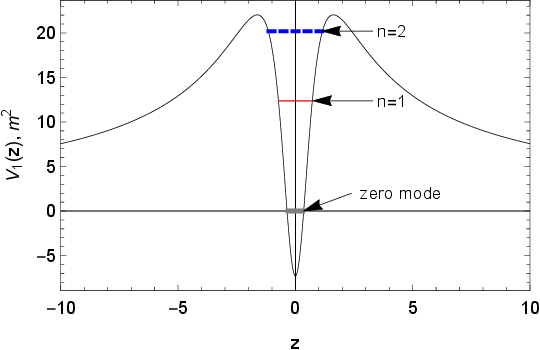}}
\hspace{0.5cm}
\subfigure[$t_1=25,t_2=0.3$.]{\label{FigResPVec}
\includegraphics[width= 0.35\textwidth]{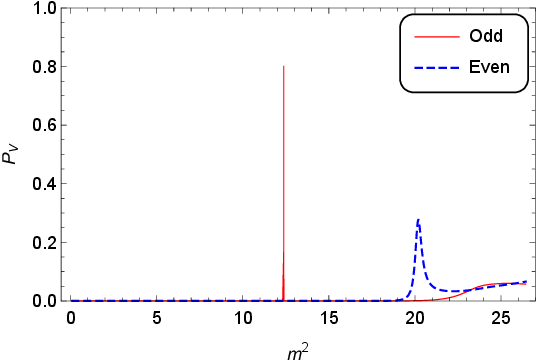}}
\subfigure[$t_1=20,t_2=0.4$.]{\label{2FigResSpecVec}
\includegraphics[width= 0.35\textwidth]{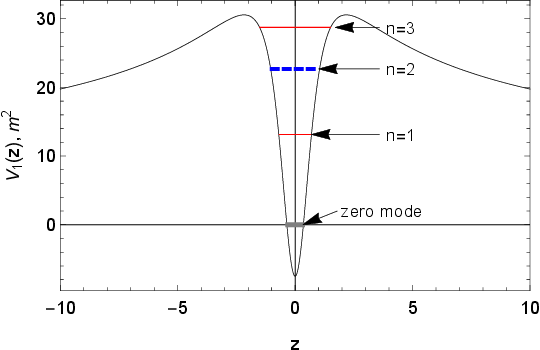}}
\hspace{0.5cm}
\subfigure[$t_1=20,t_2=0.4$.]{\label{2FigResPVec}
\includegraphics[width= 0.35\textwidth]{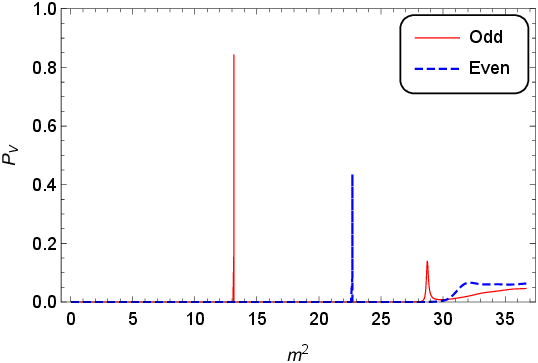}}
\subfigure[$t_1=25,t_2=0.4$.]{\label{3FigResSpecVec}
\includegraphics[width= 0.35\textwidth]{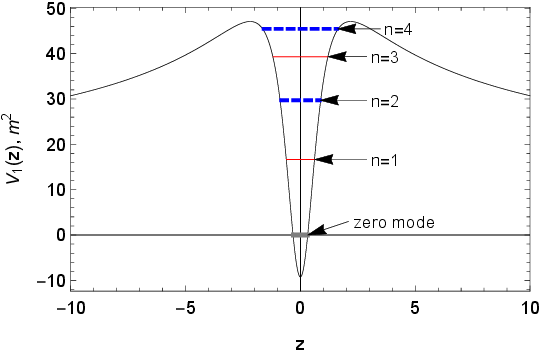}}
\hspace{0.5cm}
\subfigure[$t_1=25,t_2=0.4$.]{\label{3FigResPVec}
\includegraphics[width= 0.35\textwidth]{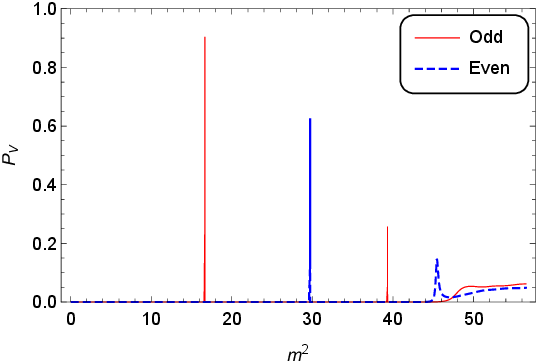}}
\end{center}\vskip -5mm
\caption{For $\mathcal{M}_4$ brane case, the mass spectra, the effective potential $V_1(z)$, and corresponding
         relative probability $P_V$ with the parameter sets: $t_1=25,t_2=0.3$; $t_1=20,t_2=0.4$;
         and $t_1=25,t_2=0.4$. $V_1(z)$ for the black line, the zero mode for the grey line,
         the even parity resonant KK modes for the blue lines, and the odd parity resonant KK modes for the
         red lines. The parameters are set as $b=2$ and $c=1$.}
 \label{FigResSpec-PVec}
\end{figure}

With the method mentioned above, the resonant KK modes can be solved from Eq. (\ref{SchroVec})
through numerical methods. Table \ref{tableRMVolVec} showcases the mass, width, and lifetime of
the resonant KK modes, considering the parameters sets: $t_1=25,t_2=0.3$; $t_1=20,t_2=0.4$; and
$t_1=25,t_2=0.4$. Notably, there is an observed increase in the number of resonant KK modes with
increasing parameters $t_1$ and $t_2$. Figure \ref{FigResSpec-PVec} presents the profiles of the
relative probability $P_V$ of the resonant modes. The corresponding mass spectra with the effective
potentials are also illustrated in Fig. \ref{FigResSpec-PVec}.

\begin{figure} 
\begin{center}
\subfigure[$\rho_1$.]{\label{fig_Odd1VolVec}
\includegraphics[width = 0.23\textwidth]{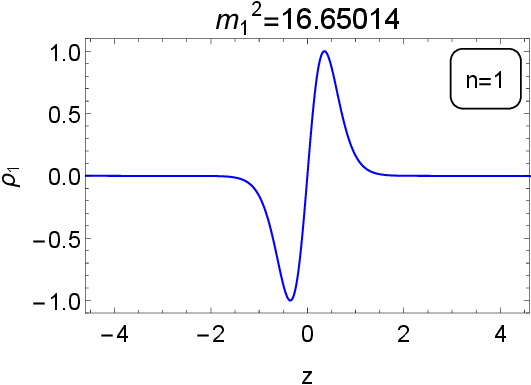}}
\subfigure[$\rho_2$.]{\label{fig_Even1VolVec}
\includegraphics[width = 0.23\textwidth]{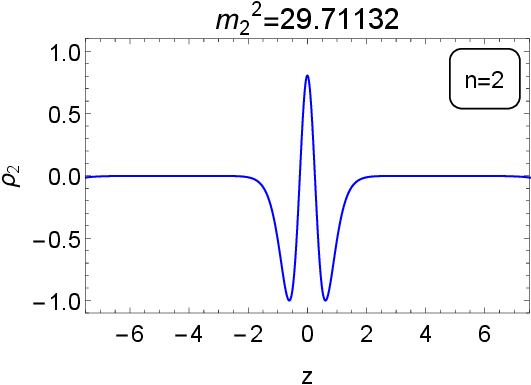}}
\subfigure[$\rho_3$.]{\label{fig_Odd2VolVec}
\includegraphics[width = 0.23\textwidth]{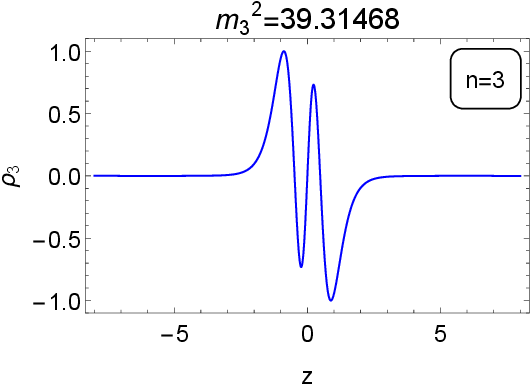}}
\subfigure[$\rho_4$.]{\label{fig_Even2VolVec}
\includegraphics[width = 0.23\textwidth]{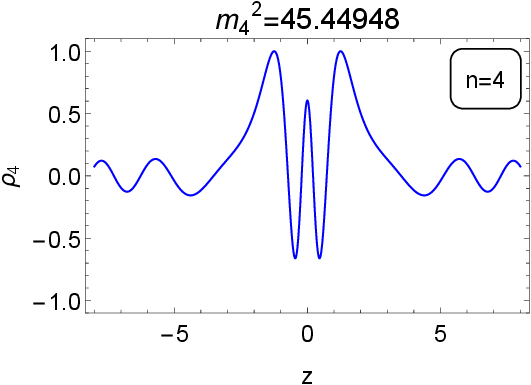}}
\end{center}\vskip -2mm
\caption{For $\mathcal{M}_4$ brane case, the shapes of resonant KK modes $\rho(z)$ for vectors
         with different $m^{2}$. The parameters are set as $b=2,c=1,
         t_1=25$ and $t_{2}=0.4$.}
 \label{FigResVec}
\end{figure}

In the mass spectra of the vector KK modes, the ground state is the zero mode (bound state),
and all the massive KK modes are resonant KK modes. Table \ref{tableRMVolVec} and Fig. \ref{3FigResPVec}
reveal the existence of four resonant KK modes for parameters $t_1=25$ and $t_2=0.4$, all of which
are graphically displayed in Fig. \ref{FigResVec}. Therefore, in the case of $0<t_2<b/4$, the vector
zero mode can be localized on the brane, and the massive KK modes can be quasi-localized on the
brane.

\begin{table}[tbp]
\centering
\begin{tabular}{|c|c|c|c|c|c|c|c|}
    \hline
    $t_1$               & $t_2$                 & $V^{\text{max}}_{1}$  & $n$         & $m^2$ &
    $m$                 & $\Gamma$              & $\tau$
    \\
    \hline
    $25$                & $0.3$                 & $22.0287$             & $1$         & $12.3803$ &
    $3.5186$            & $2.320\times10^{-5}$  & $4.309\times10^4$
    \\
                        &                       &                       & $2$         & $20.2005$ &
    $4.4945$            & $0.0530$              & $18.8813$
    \\
    \hline
    $20$                & $0.4$                 & $30.5626$             & $1$         & $13.1445$ &
    $3.6255$            & $4.265\times10^{-10}$ & $2.345\times10^9$
    \\
                        &                       &                       & $2$         & $22.6922$ &
    $4.7636$            & $5.648\times10^{-6}$  & $1.770\times10^5$
    \\
                        &                       &                       & $3$         & $28.7377$ &
    $5.3608$            & $0.0187$              & $53.3974$
    \\
    \hline
    $25$                & $0.4$                 & $47.0768$             & $1$         & $16.6501$ &
    $4.0805$            & $1.170\times10^{-13}$ & $8.530\times10^{12}$
    \\
                        &                       &                       & $2$         & $29.7113$ &
    $5.4508$            & $2.075\times10^{-9}$  & $4.820\times10^8$
    \\
                        &                       &                       & $3$         & $39.3147$ &
    $6.2701$            & $6.527\times10^{-5}$  & $1.532\times10^4$
    \\
                        &                       &                       & $4$         & $45.4495$ &
    $6.7417$            & $0.0292$              & $34.2148$
    \\
    \hline
\end{tabular}
\caption{For $\mathcal{M}_4$ brane case, the mass, width, and lifetime of resonant KK modes of the
        vectors. The parameters are set as $b=2$ and $c=1$. }
    \label{tableRMVolVec}
\end{table}

\begin{figure} 
\begin{center}
\includegraphics[width= 0.49\textwidth]{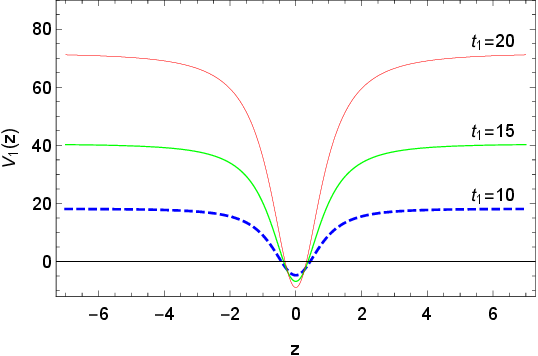}
\caption{For $\mathcal{M}_4$ brane case, the shapes of the effective potential $V_1(z)$ of the $U(1)$ gauge vector
        field with the parameter $t_1=10,15$, and $20$. The other parameters are set as $b=2,c=1$, and $t_2=0.5$.}
\label{figSpecPTVec}
\end{center}
\end{figure}

In the case where $t_2=0.5=b/4$, the effective potential adopts the P\"{o}schl-Teller potential
shape. A finite number of massive KK modes can be localized on the brane. Figure \ref{figSpecPTVec}
illustrates the shapes of the effective potentials $V_1(z)$, computed using numerical methods with
parameters $t_1=10,15,20$, and $t_2=0.5$. The figure depicts a series of potential wells located
at the brane position, with increasing depth correlated to larger value of $t_1$. Employing numerical
methods, we solved for the localized KK modes and observed that their number also rises with the
parameter $t_1$. Specifically, the corresponding mass spectra are detailed as follows:
\begin{eqnarray}
  m_n^2=&\hspace{-4.75cm}\{0,7.91,13.02,15.98,17.46\},                            &\text{for}~t_1=10,    \label{SpecPTVec1} \\
  m_n^2=&\hspace{-0.4cm}\{0,12.15,21.43,28.24,33.03,36.23,38.25,39.44,40.08\},   &\text{for}~t_1=15,    \label{SpecPTVec2} \\
  m_n^2=&\{0,16.38,29.86,40.73,49.30,55.91,60.87,64.51,67.10,~\ ~\               &\nonumber         \\
        &\hspace{-4.70cm}68.88,70.08,74.84,71.31\},                               &\text{for}~t_1=20.     \label{SpecPTVec3}
\end{eqnarray}
So in the case of $t_2=b/4$, the vector zero mode is localized on the brane. A finite number of massive
KK modes can also be localized on the brane, with the number increases alongside the parameter $t_1$.

\begin{figure} 
\begin{center}
\includegraphics[width= 0.49\textwidth]{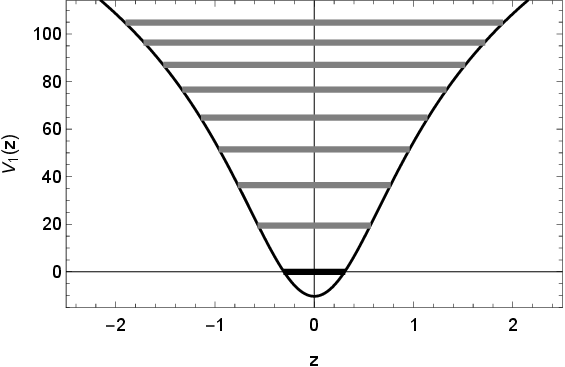}
\caption{For $\mathcal{M}_4$ brane case, the shape of the effective potential $V_1(z)$; the thick black line
        corresponds to $m_0^2=0$ and the first eight massive levels ($1\leq n\leq8$)
        of the $m_n^2$ spectrum are given by the grey lines. The parameter are
        set as $b=2,c=1,t_1=20$ and $t_2=0.6$.}
\label{figSpecIDWVec}
\end{center}
\end{figure}

Lastly, in the case where $t_2$ equals $0.6$, exceeding $b/4$, the effective potential diverges towards
positive infinity when far away from the brane, resulting in all the vector KK modes existing as bound
states. For example, the infinite discrete spectra of mass are partly indicated in Fig. (\ref{figSpecIDWVec})
with specific values of parameters.

Therefore, by considering the coupling function, the vector zero mode can be localized on the brane, and the
massive KK modes exhibit distinct behaviors, being either localized or quasi-localized on the brane.

\vspace{1cm}
\subsubsection{Kalb-Ramond fields} \label{flatKRfield}

In this section, we will investigate the localization of KK modes for the KR fields ($q=2$) with the
5D RSII-like model (\ref{Warped factor}). It is known that a  KR field is an antisymmetrical
tensor field with higher spins proposed in string theory.
In 4D spacetime, the KR field is equivalent to the scalar field by a duality, while
in higher-dimensional spacetime, it represents new particles.

Considering the coupling function $F(R)$, we assume the 5D action for a free KR field $B_{NL}$ as
\begin{eqnarray} \label{KRaction}
 S_2=-\int d^5x\sqrt{-g}F(R)H_{MNL}H^{MNL},
\end{eqnarray}
where $H_{MNL}=\partial_{[M}B_{NL]}$ is the field strength of the KR field. With the flat metric
(\ref{metricMin}), the equations of motion can be derived from the above action as follows:
\begin{eqnarray}
 \partial_{\mu}(\sqrt{-g}F(R)H^{\mu\nu\gamma})+\partial_z(\sqrt{-g}F(R)H^{z\nu\gamma})&=& 0,   \label{KREoM1}    \\
                                             \partial_{\mu}(\sqrt{-g}F(R)H^{\mu\nu z})&=& 0.   \label{KREoM2}
\end{eqnarray}
By utilizing the gauge condition $B_{\mu5}=0$ and implementing the KK decomposition
\begin{eqnarray} \label{KRDecom}
 B_{\mu\nu}(x^{\mu},z)=\sum_n\hat B^{(n)}_{\mu\nu}(x^{\mu})U^{(n)}(z)e^{\frac12A}F(R)^{-\frac12},
\end{eqnarray}
we can get the Schr\"{o}dinger-like equation
\begin{eqnarray} \label{KRSchro}
 (-\partial^2_z+V_{\text{KR}}(z))U^{(n)}(z)=m^2_nU^{(n)}(z),
\end{eqnarray}
where $m_n$ are the masses of the KK modes $U^{(n)}$ for the KR field, and the effective potential
$V_{\text{KR}}(z)$ is
\begin{eqnarray} \label{KRVy}
 V_{\text{KR}}(z)=-\frac12A''+\frac14A'^2-\frac12\frac{A'F'(R)}{F(R)}+\frac{F''(R)}{2F(R)}-\frac{F'^2(R)}{4F^2(R)}.
\end{eqnarray}

By requiring the orthonormality conditions
\begin{eqnarray} \label{KRortho}
 \int U^{(m)}U^{(n)}dz=\delta_{mn},
\end{eqnarray}
the 5D action (\ref{KRaction}) reduces to the 4D effective one:
\begin{eqnarray} \label{KR4Daction}
 S_{\text{eff}}= -\sum_n\int d^4x\sqrt{-g}(\hat h^{(n)}_{\mu\nu\gamma}\hat h^{(n)\mu\nu\gamma}
                 +\frac13m_n^2\hat B^{(n)}_{\mu\nu}\hat B^{(n)\mu\nu}),
\end{eqnarray}
where $\hat h^{(n)}_{\mu\nu\gamma}=\partial_{[\mu}\hat B_{\nu\lambda]}$ is the 4D field strength.
The localization of the KK modes need the conditions Eq. (\ref{KRortho}) to be satisfied.

From the Schr\"{o}dinger-like equation (\ref{KRSchro}), the zero mode with $m_0^2=0$ can be solved as
\begin{eqnarray} \label{KRzeroSol}
 U_0(z)=N_2e^{-\frac12A}(F(R))^{\frac12}
\end{eqnarray}
with $N_2$ the normalization constant. In light of the coordinate transformation (\ref{coordinate trans}),
the normalization condition of the zero mode reads
\begin{eqnarray} \label{KRzeroNorm}
 \int U_0^2dz&=& N_2^2\int e^{-A}F(R)dz       \nonumber    \\
      &=& N_2^2\int e^{-2A}F(R)dy=1.
\end{eqnarray}

In the context of minimal coupling where function $F(R)\equiv1$, based on the brane model
(\ref{Warped factor}), the above normalization condition (\ref{KRzeroNorm}) cannot be met due to the
divergence of the integral within it:
\begin{eqnarray} \label{KRzeroNormMinC}
 \int U_0^2dz &=& N_2^2\int e^{-2A}dy            \nonumber   \\
            &=& N_2^2\int (\cosh(cy))^bdy\rightarrow\infty,
\end{eqnarray}
so the zero mode cannot be localized on the thick brane. In addition for this case, based on the brane
model (\ref{Warped factor}) and the coordinate transformation (\ref{coordinate trans}), the effective
potential (\ref{KRSchro}) can be expressed in terms of coordinate $y$ as
\begin{eqnarray} \label{KRVyMC}
 V_{\text{KR}}(z(y))&=& -\frac14A'^2e^{2A}-\frac12A''e^{2A}           \nonumber     \\
   &=&-\frac{1}{16}bc^2(\text{sech}(cy))^{b+2}(b\sinh^2(cy)-4).
\end{eqnarray}
It can be seen that this potential tends toward zero when far away from the brane, implying no massive KK
modes trapped on the brane. Consequently, in the case of minimal coupling, neither the zero mode nor the
massive KK modes can be localized on the thick $\mathcal{M}_4$ brane.

On the contrary, with considering the coupling function $F(R)$, the localization of the KK modes becomes
feasible. For the zero mode, by substituting the warp factor (\ref{Warped factor}) and the function $F(R)$
(\ref{FR5}) into the normalization condition (\ref{KRzeroNorm}), the asymptotic behaviors of the integrand
in Eq. (\ref{KRzeroNorm}) as $y\rightarrow+\infty$ are identified as
\begin{eqnarray} \label{AsymKRzeroNorm}
 e^{-2A}F(R) &\rightarrow& 2^{-b}e^{t_1+bcy-(\frac54)^{t_2}t_1(\frac{b}{4+5b})^{t_2}e^{2t_2cy}}   \nonumber    \\
             &\rightarrow&
    \left\{
      \begin{array}{ll}
        0          \hspace{0.6cm}  & t_2>0; \\
        +\infty    \hspace{0.6cm}  & t_2<0.
      \end{array}
    \right.
\end{eqnarray}
Therefore, if parameter $t_2>0$, the zero mode of the KR field is normalizable, and can be localized on the
thick $\mathcal{M}_4$ brane. If $t_2<0$, the zero mode cannot be normalized.

\begin{figure} 
\begin{center}
\subfigure[$V_{\text{KR}}(z)$.]{\label{FigeffVKR}
\includegraphics[width= 0.48\textwidth]{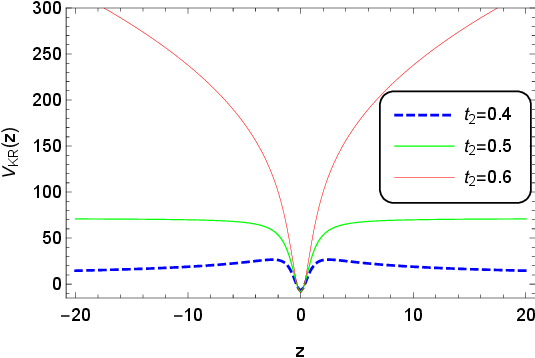}}
\subfigure[$U_0(z)$.]{\label{FigZMKR}
\includegraphics[width= 0.48\textwidth]{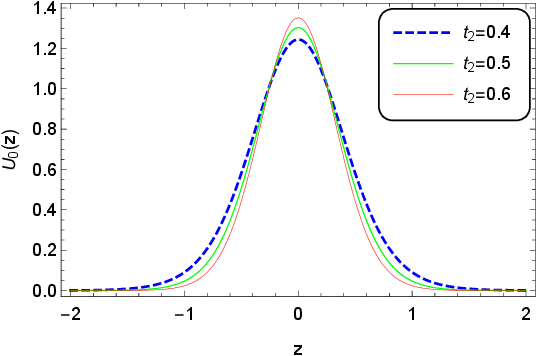}}
\end{center}\vskip -5mm
\caption{For $\mathcal{M}_4$ brane case, the effective potentials $V_{\text{KR}}(z)$ in (a), and the shapes of the zero mode $U_0(z)$
         for the Kalb-Ramond (KR) field  in (b). The parameters are set as $b=2,c=1,t_1=20$, and
         $t_{2}=0.4,0.5,0.6$.}
 \label{FigZM-EPKR}
\end{figure}

Concerning the massive KK modes, employing the coordinate transformation (\ref{coordinate trans}), we
can express the effective potential $V_{\text{KR}}(z)$ (\ref{KRVy}) with respect to coordinate $y$ as
\begin{eqnarray} \label{KRVzy}
  V_{\text{KR}}(z(y))=-\frac14 A'^2e^{2A}-\frac12 A''e^{2A}+\frac{F''(R)e^{2A}}{2F(R)}-\frac{F'^2(R)e^{2A}}{4F^2(R)}.
\end{eqnarray}
Substituting Eqs. (\ref{Warped factor}) and (\ref{FR5}) into the above expression, we can get
\begin{eqnarray} \label{KRVzy1}
  V_{\text{KR}}(z(y))
    &=& \frac{c^2}{16}(\text{sech}(cy))^{b+2}\bigg(\frac{(5b+4)^2\text{sech}^4(cy)}{b^2}\bigg)^{-t_2}          \nonumber   \\
    & & \times\bigg[16\times25^{t_2}t_1^2t_2^2\sinh^2(cy)-b\bigg(\frac{(5b+4)^2\text{sech}^4(cy)}{b^2}\bigg)^{t_2}(-4+b\sinh^2(cy))  \nonumber \\
    & & -16\times5^{t_2}t_1t_2\bigg(\frac{(5b+4)^2\text{sech}^4(cy)}{b^2}\bigg)^{t_2/2}(1+2t_2\sinh^2(cy))\bigg].
\end{eqnarray}
From this expression, the asymptotic behaviors of the effective potential as $y\rightarrow+\infty$
can be obtained:
\begin{eqnarray} \label{AsymEffPotenKR}
  V_{\text{KR}}(z(y\rightarrow+\infty))\rightarrow \left\{
    \begin{array}{llcll}
      +\infty                       & \hspace{0.8cm}  t_2   > b/4,     \\
      \mathrm{C}_{\text{KR}}        & \hspace{0.8cm}  t_2   = b/4,     \\
      0                             & \hspace{0.8cm}  0<t_2 < b/4
\end{array} \right.
\end{eqnarray}
with the limit
\begin{eqnarray} \label{LmtKR}
  \mathrm{C}_{\text{KR}}=\frac{1}{16}b^2c^2t_1^2\bigg(\frac{5b}{5b+4}\bigg)^{b/2}.
\end{eqnarray}
Likewise, in the case of the KR field, a positive parameter $t_2$ is necessary to realize the
localization of the zero mode. With focusing on the cases where $t_2$ spans around the critical
value $b/4$, the effective potential $V_{\text{KR}}(z)$ and zero mode $U_0(z)$ are illustrated
in Fig. \ref{FigZM-EPKR} using numerical methods with parameters: $b=2,c=1,t_1=20$, and
$t_2=0.4,0.5,0.6$.

In the case where $t_2=0.4<b/4$, the effective potential converges to zero as $z\rightarrow\pm\infty$.
There is no localized massive modes on the thick brane, but the resonant KK modes could exist. Employing
the approach presented in Refs. \cite{YXLiuPRD0980,YXLiuPRD0980-2}, we define the relative probability
function of a resonance for the KR field as
\begin{eqnarray} \label{RPResKR}
   P_{\text{KR}}(m^2)=\frac{\int^{z_b}_{-z_b}|U(z)|^2dz}{\int^{z_{max}}_{-z_{max}}|U(z)|^2dz},
\end{eqnarray}
where $2z_b$ represents approximately the width of the thick brane, and $z_{max}$ is set as $10z_b$.
The lifetime $\tau$ of a resonant mode is $\tau\sim\Gamma^{-1}$, where $\Gamma=\delta m$ denotes
the full width at half maximum of the resonant peak. The resonant modes can be solved from
Eq. (\ref{KRSchro}) with numerical methods. Our calculations will utilize specific parameter sets:
$t_1=25,t_2=0.3$; $t_1=20,t_2=0.4$; and $t_1=25,t_2=0.4$.

\begin{figure} 
\begin{center}
\subfigure[$t_1=25,t_2=0.3$.]{\label{FigResSpecKR}
\includegraphics[width= 0.35\textwidth]{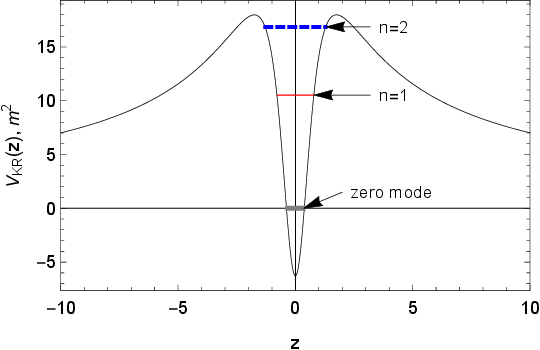}}
\hspace{0.5cm}
\subfigure[$t_1=25,t_2=0.3$.]{\label{FigResPKR}
\includegraphics[width= 0.35\textwidth]{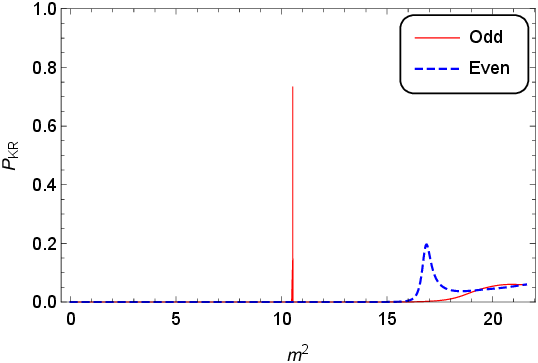}}
\subfigure[$t_1=20,t_2=0.4$.]{\label{2FigResSpecKR}
\includegraphics[width= 0.35\textwidth]{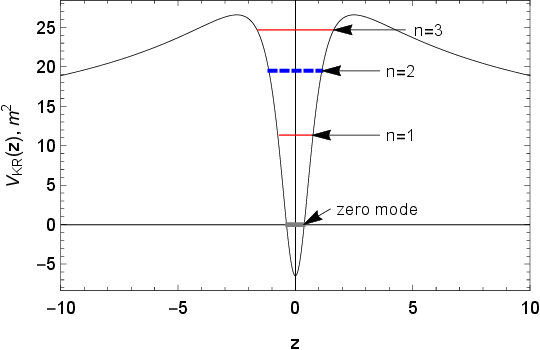}}
\hspace{0.5cm}
\subfigure[$t_1=20,t_2=0.4$.]{\label{2FigResPKR}
\includegraphics[width= 0.35\textwidth]{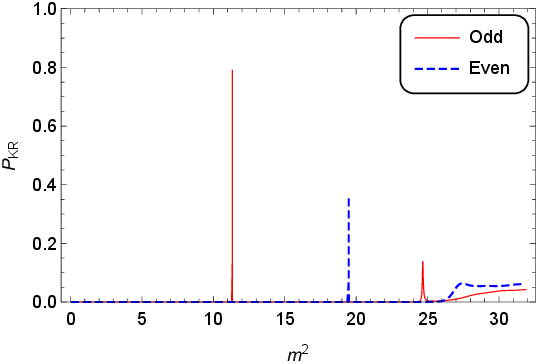}}
\subfigure[$t_1=25,t_2=0.4$.]{\label{3FigResSpecKR}
\includegraphics[width= 0.35\textwidth]{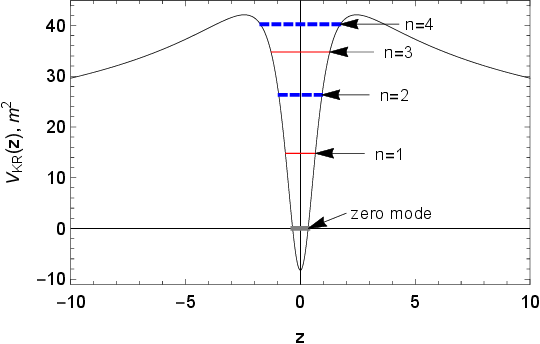}}
\hspace{0.5cm}
\subfigure[$t_1=25,t_2=0.4$.]{\label{3FigResPKR}
\includegraphics[width= 0.35\textwidth]{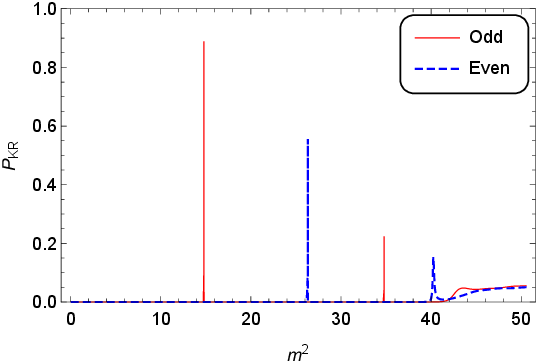}}
\end{center}\vskip -5mm
\caption{For $\mathcal{M}_4$ brane case, the mass spectra, the effective potential $V_{\text{KR}}(z)$, and corresponding
         relative probability $P_{\text{KR}}$ with parameter sets: $t_1=25,t_2=0.3$; $t_1=20,t_2=0.4$;
         and $t_1=25,t_2=0.4$. $V_{\text{KR}}(z)$ for the black line, the zero mode for the grey line,
         the even parity resonant KK modes for the blue lines, and the odd parity resonant KK modes for the red lines.
         The parameters are set as $b=2$ and $c=1$.}
 \label{FigResSpec-PKR}
\end{figure}

\begin{figure} 
\begin{center}
\subfigure[$U_1$.]{\label{fig_Odd1VolKR}
\includegraphics[width = 0.23\textwidth]{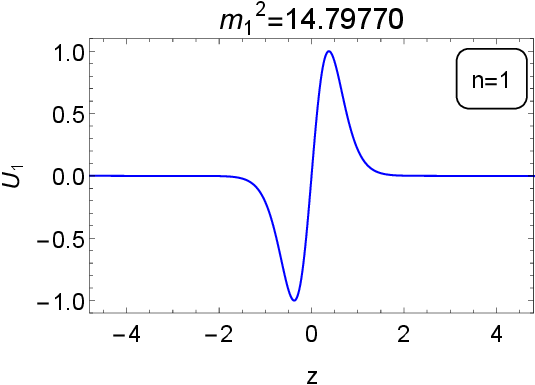}}
\subfigure[$U_2$.]{\label{fig_Even1VolKR}
\includegraphics[width = 0.23\textwidth]{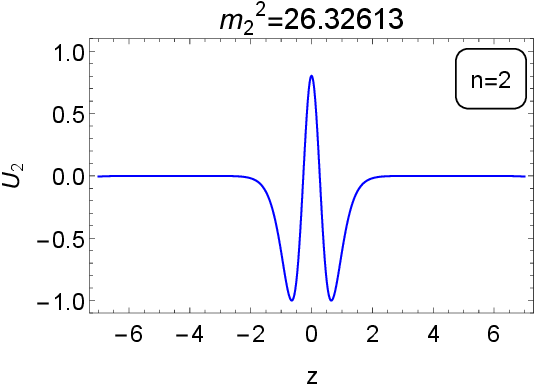}}
\subfigure[$U_3$.]{\label{fig_Odd2VolKR}
\includegraphics[width = 0.23\textwidth]{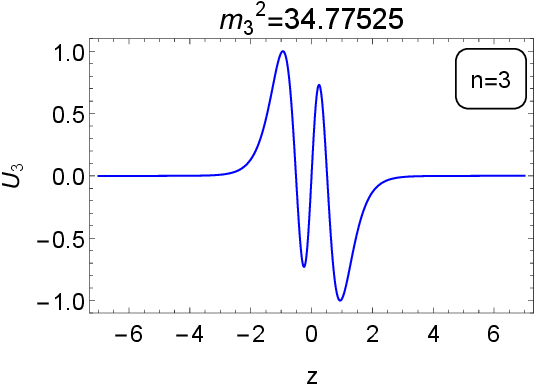}}
\subfigure[$U_4$.]{\label{fig_Even2VolKR}
\includegraphics[width = 0.23\textwidth]{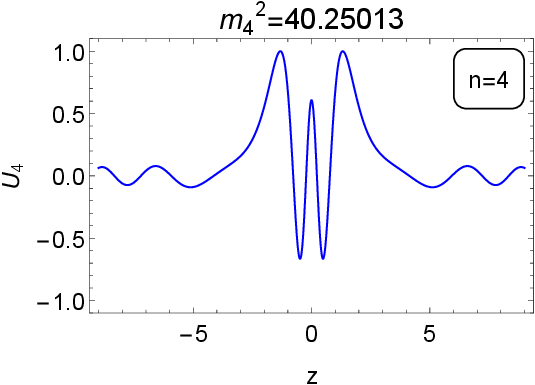}}
\end{center}\vskip -2mm
\caption{For $\mathcal{M}_4$ brane case, the shapes of resonant KK modes $U(z)$ for the KR field
         with different $m^{2}$. The parameters are set as $b=2,c=1,
         t_1=25$ and $t_{2}=0.4$.}
 \label{FigResKR}
\end{figure}

For the given parameter sets, resonant KK modes consistently exist on the thick brane. The mass,
width, and lifetime of all resonant KK modes are detailed in Table. \ref{tableRMVolKR}. It is
evident that the number of the resonant KK modes increases with the values of parameters $t_1$ and $t_2$.
Profiles of the corresponding relative probability, $P_{\text{KR}}$, are presented in Fig.
\ref{FigResSpec-PKR}. The associated
mass spectra, along with the effective potentials, are illustrated in the same figure. In this
depiction, the zero mode serves as the ground state (bound state), and all massive KK modes are resonant
KK modes. The four resonant KK modes in the specific case of $t_1=25$ and $t_2=0.4$ are depicted in
Fig. \ref{FigResKR}. Consequently, within the parameter range $0<t_2<b/4$, the zero mode for the KR
field is localized on the brane, and the massive KK modes can be quasi-localized on the brane.

\begin{table}[tbp]
\centering
\begin{tabular}{|c|c|c|c|c|c|c|c|}
    \hline
    $t_1$               & $t_2$                 & $V^{\text{max}}_{\text{KR}}$  & $n$         & $m^2$ &
    $m$                 & $\Gamma$              & $\tau$
    \\
    \hline
    $25$                & $0.3$                 & $17.9882$             & $1$         & $10.5314$ &
    $3.2452$            & $3.682\times10^{-5}$  & $2.716\times10^4$
    \\
                        &                       &                       & $2$         & $16.8650$ &
    $4.1067$            & $0.0702$              & $14.2534$
    \\
    \hline
    $20$                & $0.4$                 & $26.5798$             & $1$         & $11.3323$ &
    $3.3663$            & $1.161\times10^{-9}$  & $8.616\times10^8$
    \\
                        &                       &                       & $2$         & $19.4845$ &
    $4.4141$            & $3.604\times10^{-6}$  & $2.775\times10^5$
    \\
                        &                       &                       & $3$         & $24.6688$ &
    $4.9668$            & $0.0080$              & $124.3212$
    \\
    \hline
    $25$                & $0.4$                 & $42.1007$             & $1$         & $14.7977$ &
    $3.8468$            & $5.610\times10^{-13}$ & $1.781\times10^{12}$
    \\
                        &                       &                       & $2$         & $26.3261$ &
    $5.1309$            & $2.545\times10^{-9}$  & $3.929\times10^8$
    \\
                        &                       &                       & $3$         & $34.7753$ &
    $5.8971$            & $1.435\times10^{-5}$  & $6.968\times10^4$
    \\
                        &                       &                       & $4$         & $40.2501$ &
    $6.3443$            & $0.0152$              & $65.8030$
    \\
    \hline
\end{tabular}
\caption{For $\mathcal{M}_4$ brane case, the mass, width, and lifetime of resonant KK modes for the
         KR field. The parameters are set as $b=2$ and $c=1$. }
    \label{tableRMVolKR}
\end{table}

\begin{figure} 
\begin{center}
\includegraphics[width= 0.49\textwidth]{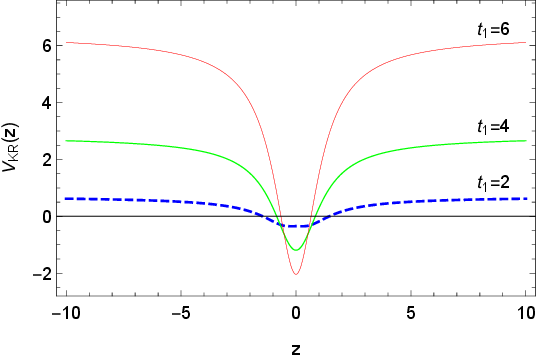}
\caption{For $\mathcal{M}_4$ brane case, the shapes of the 
        effective potential $V_{\text{KR}}(z)$
        for the parameter $t_1=2,4$, and $6$. The other parameters are set as $b=2,
        c=1$, and $t_2=0.5$.}
\label{figSpecPTKR}
\end{center}
\end{figure}

In the case where $t_2=0.5=b/4$, the effective potential tends to the limit $\text{C}_{\text{KR}}$
when far away from the brane. Hence, a finite number of localized KK modes exist. The profiles of
the effective potential $V_{\text{KR}}(z)$ are depicted in Fig. \ref{figSpecPTKR} with parameters
$t_1=2,4,6$, and $t_2=0.5$. As illustrated in the figure, the effective potentials exhibit a series
of potential wells, with the depth of these potential wells increasing alongside the parameter
$t_1$. The corresponding mass spectra of the localized KK modes  are presented below:
\begin{eqnarray}
  m_n^2=&\hspace{-2.90cm}\{0,0.27,0.62,0.66,0.69,0.71\},                    &\text{for}~t_1=2,  \label{SpecPTKR1} \\
  m_n^2=&\hspace{-0.34cm}\{0,1.66,2.30,2.57,2.68,2.74,2.78,2.81,2.84\},     &\text{for}~t_1=4,  \label{SpecPTKR2} \\
  m_n^2=&\{0,3.11,4.70,5.48,5.87,6.07,6.19,6.26,6.30,~\ ~\                  &\nonumber         \\
        &\hspace{-4.84cm}6.34,6.27,6.42\},                                  &\text{for}~t_1=6.     \label{SpecPTKR3}
\end{eqnarray}
Analysis of these mass spectra reveals a finite number of the localized massive KK modes, with the number
increasing as $t_1$ rises. Moreover, upon comparing the aforementioned mass spectra with the analogous mass spectra
for the scalar field (\ref{SpecPTSca1},\ref{SpecPTSca2},\ref{SpecPTSca3}) and the $U(1)$ gauge vector field (\ref{SpecPTVec1},\ref{SpecPTVec2},\ref{SpecPTVec3}), it is evident that when $t_2=b/4$, the number of the
localized KK modes for $q-$form field increases with the index $q$.

\begin{figure}
\begin{center}
\includegraphics[width= 0.49\textwidth]{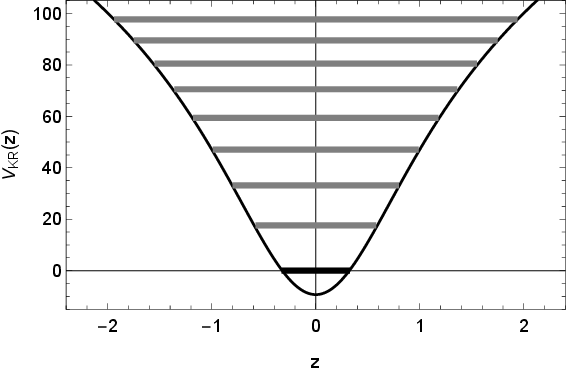}
\caption{For $\mathcal{M}_4$ brane case, the shape of the effective potential $V_{\text{KR}}(z)$; the thick black line
        corresponds to $m_0^2=0$ and the first eight massive levels ($1\leq n\leq8$)
        of the $m_n^2$ spectrum are given by the grey lines. The parameter are
        set as $b=2,c=1,t_1=20$ and $t_2=0.6$.}
\label{figSpecIDWKR}
\end{center}
\end{figure}

Finally, in the case where $t_2=0.6>b/4$, an infinitely deep well emerges, leading to the localization of all
massive KK modes on the thick brane. The mass spectra corresponding to lower localized KK modes are shown in
Fig. \ref{figSpecIDWKR} with parameters $b=2,c=1,t_1=20$, and $t_2=0.6$.

Therefore, when parameter $t_2>0$, the zero mode of KR field can be localized on the brane, and
the massive KK modes can be localized or quasi-localized on the brane.

In addition, at position $y=0$, the effective potential (\ref{KRVzy1}) becomes
\begin{eqnarray} \label{KRVzy=0}
  V_{\text{KR}}(z(y=0))=\frac14c^2\bigg[b-4t_1t_2\bigg(\frac{5b}{5b+4}\bigg)^{t_2}\bigg].
\end{eqnarray}
It can be seen that the value of potential $V_{\text{KR}}(z(y=0))$
tends toward $\frac14bc^2$ in the cases of $t_2\rightarrow 0$ or $t_2\rightarrow+\infty$. Besides,
it exhibits the following minimum:
\begin{eqnarray} \label{MiniKRVzy=0}
  V_{\text{KR}}(z(y=0))=\frac14c^2\bigg(b+\frac{4t_1}{e\ln(\frac{5b}{5b+4})}\bigg),~ \text{when}~t_2=-\frac{1}{\ln(\frac{5b}{5b+4})}.
\end{eqnarray}
At this point, when $t_1>-\frac14be\ln(\frac{5b}{5b+4})$, a negative potential well could form
within the effective potential at $y=0$. Then, if the negative potential well emerges
under the condition of $0<t_2<b/4$, resulting in a volcanic potential, the resonant KK modes
could exist on the thick brane, as indicated in Fig. \ref{FigResSpec-PKR}.

\begin{figure} 
\begin{center}
\subfigure[$V_{\text{KR}}(z)$.]{\label{FigeffVOriKR}
\includegraphics[width= 0.48\textwidth]{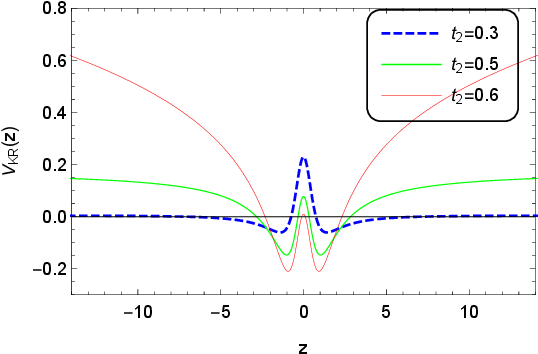}}
\subfigure[$U_0(z)$.]{\label{FigZMOriKR}
\includegraphics[width= 0.48\textwidth]{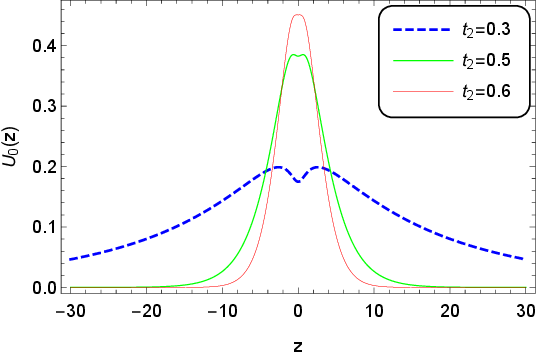}}
\end{center}\vskip -5mm
\caption{For $\mathcal{M}_4$ brane case, the effective potentials $V_{\text{KR}}(z)$ in (a), and the
         shapes of the zero mode $U_0(z)$ for the KR field in (b). The parameters are set as $b=2,c=1,
         t_1=1$,and $t_{2}=0.3,0.5,0.6$.}
 \label{FigZMEffOriKR}
\end{figure}

On the other hand, for the case of $0<t_1<-\frac14be\ln(\frac{5b}{5b+4})$, the effective potential
consistently maintains positive at the origin of the extra dimension. Concerning the positivity of
the effective potential $V_{\text{KR}}(z(y=0))$, we observe the emergence of positive barriers at
the brane position. These barriers could contribute to the formation of a local minimum shaping the
zero mode at the origin of the extra dimension, as depicted in Fig. \ref{FigZMEffOriKR}. However,
the value of potential $V_{\text{KR}}(z(y=0))$ always be lower than value $\frac14bc^2$, and the
zero mode will not approach zero at the origin of the extra dimension. Additionally, in
Fig. \ref{FigeffVOriKR}, as parameter $t_2$ increases, the potential $V_{\text{KR}}(z\rightarrow\pm\infty)$
tends to zero, limit $500/7$ and positive infinity in sequence.

\subsection{dS$_4$ brane}   \label{dS}

For the dS$_4$ brane case, the brane spacetime is warped with a positive scalar curvature. We consider
the following brane model \cite{AWangPRD}:
\begin{eqnarray}
  A(z) &=& -\delta \ln\big[\cosh\big(\frac{Hz}{\delta}\big)\big],       \label{dSbrane}    \\
  \varphi(z) &=& \sqrt{3\delta(1-\delta)}\arcsin\big[\tanh(\frac{Hz}{\delta})\big],
\end{eqnarray}
where $\delta$ is a constant satisfying $0<\delta\leq 2/3$, $H$ represents the dS parameter, and the 4D
cosmological constant is denoted as $\Lambda_4=3H^2$. The stability of this brane model has been demonstrated
in Refs. \cite{AWangPRD,HGuoEL1297,guo2023scalar}.

The bulk scalar curvature can be obtained as
\begin{eqnarray} \label{curvaturedS}
  R(z)=\frac{4H^2(3\delta+2)}{\delta}(\cosh(\frac{Hz}{\delta}))^{2\delta-2}.
\end{eqnarray}
From this expression, we can see that the bulk spacetime is asymptotically flat, and the brane model
(\ref{dSbrane}) is regular with no singularity.

Furthermore, we consider the function $F(R)$ in the form of
\cite{guo2023scalar}
\begin{eqnarray} \label{dSFR}
  F(R)=e^{t_1(1-e^{t_2R})},
\end{eqnarray}
where parameters $t_1$ and $t_2$ are positive. Then, for the zero mode of the $q-$form field (\ref{zero mode}),
there is
\begin{eqnarray} \label{ZMdS}
  \tilde U(z)= \big[\cosh(\frac{Hz}{\delta})\big]^{(q-\frac32)\delta}
               \sqrt{e^{t_1\big[1-e^{\frac{4H^2t_2(3\delta+2)}{\delta}(\cosh(\frac{Hz}{\delta}))^{2\delta-2}}\big]}}.
\end{eqnarray}
In light of the localization condition (\ref{CondLocaZM}), we can get
\begin{eqnarray} \label{CondLocaZMdS}
 \int \tilde U_0^2 dz = \int \big[\cosh(\frac{Hz}{\delta})\big]^{(2q-3)\delta}
               e^{t_1\big[1-e^{\frac{4H^2t_2(3\delta+2)}{\delta}(\cosh(\frac{Hz}{\delta}))^{2\delta-2}}\big]}dz.
\end{eqnarray}
It can be seen that if the index $q<3/2$, the zero mode can be normalized.

Regarding this coupling mechanism, the localization of the scalar KK modes has been discussed in
Ref. \cite{guo2023scalar} within the same dS$_4$ brane model. It is found that introducing the
function $F(R)$ will not change the asymptotical behaviors of the effective potential for the KK
modes when far away from the brane, while the shape of the effective potential varies at finite
position. Then, the scalar zero mode is localized on the thick brane, maybe with a split probability
density, and the massive modes could be quasi-localized at the brane position. Therefore, we will
investigate the localization of other $q-$form fields here.

\subsubsection{$U(1)$ gauge vector fields}  \label{dSvector}

For the $U(1)$ gauge vector fields, we will directly employ the corresponding solution of the zero
mode (\ref{zero mode}) and the effective potential (\ref{eff-potential}). Then, the vector zero mode
can be expressed as
\begin{eqnarray}  \label{ZMvecdS}
  \rho_0(z) =\cosh^{-\frac{\delta}{2}}(\frac{Hz}{\delta})\sqrt{e^{t_1-t_1
             e^{\frac{4H^2t_2(3\delta+2)}{\delta}}(\cosh(\frac{Hz}{\delta}))^{2\delta-2}}}.
\end{eqnarray}
The normalization of this zero mode requires
\begin{eqnarray}  \label{NormZMvecdS}
  \int\rho_0^2dz =\int\cosh^{-\delta}(\frac{Hz}{\delta})e^{t_1\big[1-
             e^{\frac{4H^2t_2(3\delta+2)}{\delta}}(\cosh(\frac{Hz}{\delta}))^{2\delta-2}\big]}dz<\infty.
\end{eqnarray}
We can see that as $0<\delta\leq 2/3$, the vector zero mode is always normalizable, and can be localized
on the thick dS$_4$ brane.

\begin{figure} 
\begin{center}
\subfigure[$V_1(z)$.]{\label{FigEffPotVecdS}
\includegraphics[width= 0.48\textwidth]{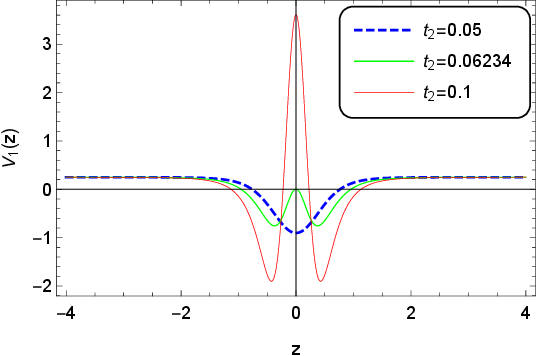}}
\subfigure[$\rho_0(z)$.]{\label{FigZMVecdS}
\includegraphics[width= 0.48\textwidth]{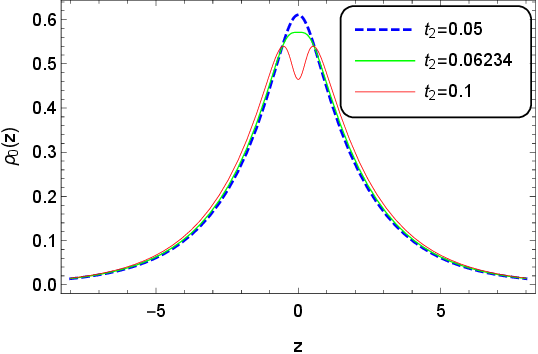}}
\end{center}\vskip -5mm
\caption{For the dS$_4$ brane case, the effective potentials $V_1(z)$ in (a), and the shapes of the zero mode $\rho_0(z)$ for
         the $U(1)$ gauge vector field in (b). The parameters are set as $\delta=0.5,H=1,t_1=0.05$,
         and $t_{2}=0.02,0.06234,0.1$.}
 \label{FigZMEffPotVecdS}
\end{figure}

For the effective potential, since the function $F(R)$ tends to value $1$ as $z\rightarrow \pm\infty$,
the coupling returns to the minimal coupling. Thus, the effective potential exhibits the following
asymptotic behavior:
\begin{eqnarray}  \label{AsymEffPotvecdS}
  V_1(z\rightarrow \pm\infty)&\rightarrow& \frac12A''+\frac14A'^2                              \nonumber   \\
                           &=& \frac{H^2}{4\delta}\big[\delta-(\delta+2)\text{sech}^2(\frac{Hz}{\delta})\big].
\end{eqnarray}
Therefore, the effective potential tends towards $H^2/4$ as $z\rightarrow \pm\infty$. For this
effective potential, no localized massive vector KK modes are found on the brane. However, there
are more abundant informations at finite positions of the extra dimension. We show the profiles
of the vector zero mode and the effective potential in Fig. \ref{FigZMEffPotVecdS} with specifical
values of parameters. It can be seen that as parameter $t_2$ increases, a potential barrier
emerges at brane position, and then the maximum shaping the zero mode at origin splits into two.

\begin{figure} 
\begin{center}
\subfigure[$V_1(z)$.]{\label{FigOriEffPotVecdS}
\includegraphics[width= 0.48\textwidth]{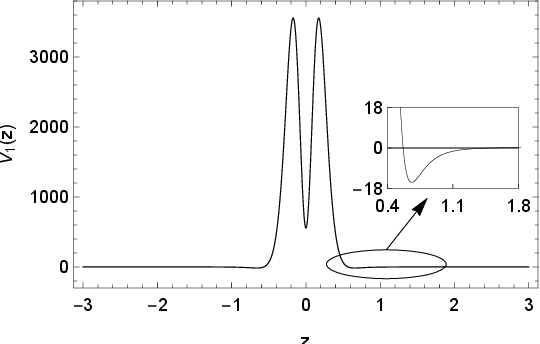}}
\subfigure[$\rho_0(z)$.]{\label{FigOriZMVecdS}
\includegraphics[width= 0.48\textwidth]{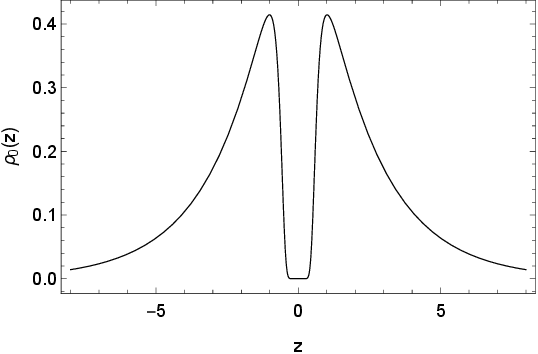}}
\end{center}\vskip -5mm
\caption{For the dS$_4$ brane case, the effective potentials $V_1(z)$ in (a), and the shapes of the
         zero mode $\rho_0(z)$ for the $U(1)$ gauge vector field in (b). The parameters are set as
         $\delta=0.5,H=1,t_1=0.05$, and $t_{2}=0.24$.}
 \label{FigOriZMEffPotVecdS}
\end{figure}

Furthermore, the potential barrier will turn into a potential well with a positive lower boundary
as parameters $t_1$ and $t_2$ rise. For example, we depict the zero mode and the effective potential
in Fig. \ref{FigOriZMEffPotVecdS} with a larger value of $t_2$. It can be seen that a potential well
with positive lower boundary is located at the origin. Then, the zero mode is localized on both sides
of the origin of the extra dimension.

\begin{figure} 
\begin{center}
\includegraphics[width= 0.96\textwidth]{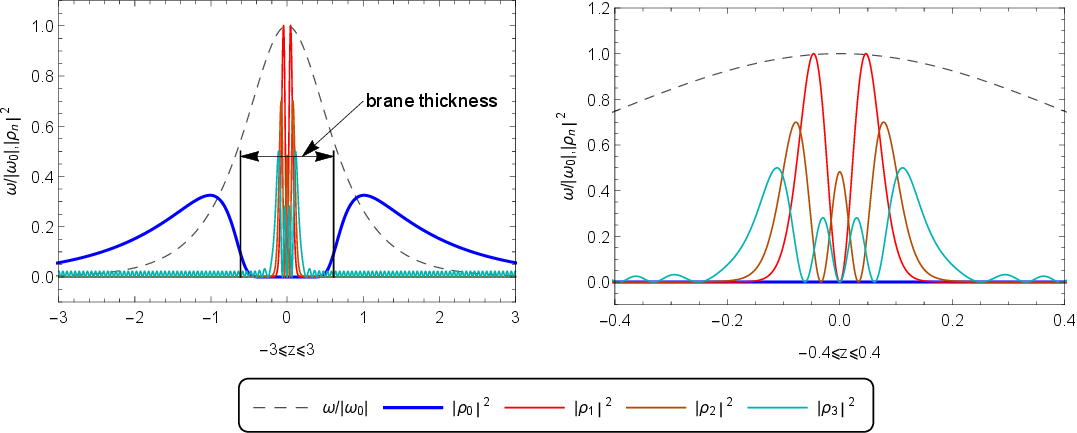}
\caption{For the dS$_4$ brane case, the relative energy density $\omega/|\omega_0|$ of the 5D dS brane
        model for the dashed line, the probability densities $|\rho_n|^2$ of the vector zero mode and the
        vector resonant modes for the colored lines. The parameters are set as $\delta=0.5,H=1,t_1=0.05$
        and $t_2=0.24$.}
\label{EDPDResoZMVecdS}
\end{center}
\end{figure}

For the potential well at brane position, this configuration can quasi-localize the massive vector
modes on the brane. We could solve for the resonant modes from the Schr\"{o}dinger-like equation.
In Fig. \ref{EDPDResoZMVecdS}, the probability densities of the vector zero mode and the resonant
modes are visually represented, along with the relative energy density $\omega/|\omega_0|$. The
brane thickness of the dS$_4$ brane is also identified in the figure. It can be observed that there
are three resonant modes existing at the brane position. The probability densities of the zero mode
and the resonant modes are concentrated in distinct regions of the extra dimension. This distribution
can be regarded as a separation of vectors along the extra dimension. For this dS$_4$ brane model
(\ref{dSbrane}), a similar phenomenon has been reported for the scalar fields \cite{guo2023scalar}.

\subsubsection{Kalb-Ramond fields}  \label{dSKR}

Concerning the KR fields, in terms of Eq. (\ref{zero mode}), the zero mode can be expressed as
\begin{eqnarray}  \label{ZMKRdS}
  U_0(z) = \cosh^{\frac{\delta}{2}}(\frac{Hz}{\delta})\sqrt{e^{t_1-t_1
           e^{\frac{4H^2t_2(3\delta+2)}{\delta}(\cosh(\frac{Hz}{\delta}))^{2\delta-2}}}}.
\end{eqnarray}
Based on the localization condition (\ref{CondLocaZM}), there is
\begin{eqnarray}  \label{NormZMKRdS}
  \int U_0^2dz = \int\cosh^{\delta}(\frac{Hz}{\delta})e^{t_1\big[1-
           e^{\frac{4H^2t_2(3\delta+2)}{\delta}(\cosh(\frac{Hz}{\delta}))^{2\delta-2}}\big]}dz.
\end{eqnarray}
As parameter $0<\delta\leq2/3$, this zero mode of the KR field cannot be normalized, so it cannot
be localized on the thick dS$_4$ brane.

\begin{figure} 
\begin{center}
\subfigure[$V_{\text{KR}}(z)$.]{\label{FigEffPotKRdS}
\includegraphics[width= 0.48\textwidth]{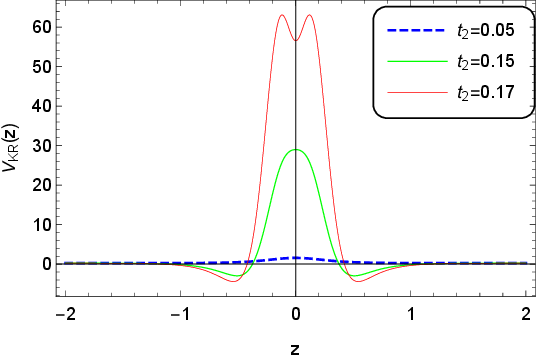}}
\subfigure[$U_0(z)$.]{\label{FigZMKRdS}
\includegraphics[width= 0.48\textwidth]{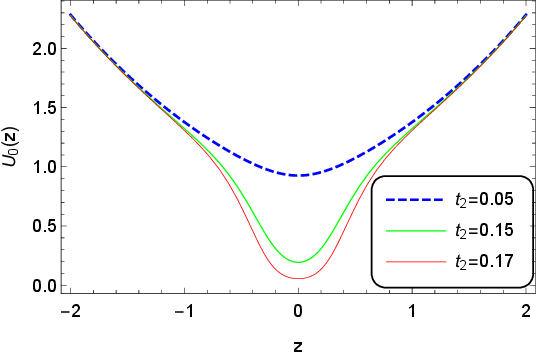}}
\end{center}\vskip -5mm
\caption{For the dS$_4$ brane case, the effective potential $V_{\text{KR}}(z)$ and the zero mode $U_0(z)$
         of the KR fields. The parameters are set as $\delta=0.5,H=1,t_1=0.05$, and $t_2=0.05,0.15,0.17$.}
 \label{UniFigEffPotKRdS}
\end{figure}

Then, for the massive KK modes, the coupling function $F(R)$ (\ref{dSFR}) tends to value $1$ as
$z\rightarrow \pm\infty$, the corresponding effective potential of KK modes (\ref{eff-potential})
exhibits the following asymptotic behaviors:
\begin{eqnarray}  \label{AsymEffPotKRdS}
  V_{\text{KR}}(z\rightarrow \pm\infty)&\rightarrow& -\frac12A''+\frac14A'^2                              \nonumber   \\
                           &=& \frac{H^2}{4\delta}\big[\delta-(\delta-2)\text{sech}^2(\frac{Hz}{\delta})\big].
\end{eqnarray}
From this expression, it can be seen that the effective potential tends to $H^2/4$ as
$z\rightarrow \pm\infty$. In Fig. \ref{UniFigEffPotKRdS}, we plot this potential $V_{\text{KR}}(z)$
alongside the zero mode $U_0(z)$ for specifical parameter values. The zero mode remains delocalized,
and there exists a continuum spectrum of massive KK modes with $m^2>H^2/4$. As the parameter $t_2$
increases, a potential barrier emerges at the brane position, accompanied by two wells on either
side of the origin. However, these wells do not alter the delocalized nature of the zero mode.

\subsection{AdS$_4$ brane}   \label{AdS}

Lastly for the AdS$_4$ brane case, the scalar curvature of the brane spacetime remains negative. The brane model that
we consider is \cite{AWangPRD}
\begin{eqnarray}
  A(z) &=& -\delta \ln\big|\cos(\frac{H}{\delta}z)\big|,           \label{AdSbrane}   \\
  \varphi(z) &=& \sqrt{3\delta(\delta-1)}\text{arcsinh}\big[\tan(\frac{H}{\delta}z)\big],
\end{eqnarray}
where parameter $\delta>1$, and the 4D cosmological constant $\Lambda_4=-3H^2$. The domain of the extra dimension is
$-z_b\leq z\leq z_b$ with $z_b=|\frac{\delta\pi}{2H}|$. The metric with this warp factor exhibits a naked singularity
at $\pm z_b$. The stability of this brane model has been demonstrated in Refs. \cite{HGuoEL1297,YXLiuPRD1184,guo2023scalar}.

For this brane model, the 5D scalar curvature reads
\begin{eqnarray} \label{curvatureAdS}
  R(z)=-\frac{4H^2(3\delta+2)}{\delta}\big[\cos(\frac{H}{\delta}z)\big]^{2\delta-2}.
\end{eqnarray}
It can be seen that as parameter $\delta>1$, the bulk scalar curvature tends to zero at the boundaries of the extra dimension.
So, the bulk spacetime is asymptotically flat, and this AdS brane model (\ref{AdSbrane}) is regular.

Then, we suggest the coupling function $F(R)$ of the form \cite{guo2023scalar}
\begin{eqnarray} \label{FRAdS}
  F(R)=e^{t_1(1-e^{-t_2R})},
\end{eqnarray}
where both parameters $t_1$ and $t_2$ are positive. At the boundaries of the extra dimension, since the bulk
scalar curvature vanishes, the function $F(R)$ tends to value $1$.

In terms of the warp factor (\ref{AdSbrane}) and the coupling function (\ref{FRAdS}), the zero mode for the
$q-$form field (\ref{zero mode}) can be expressed as
\begin{eqnarray} \label{ZMAdS}
  \tilde U_0(z)=\big[\cos(\frac{H}{\delta}z)\big]^{(q-\frac32)\delta}\sqrt{e^{t_1
                \big[1-e^{\frac{4H^2t_2(3\delta+2)}{\delta}(\cos(\frac{H}{\delta}z))^{2\delta-2}}\big]}},
\end{eqnarray}
and its normalization requires
\begin{eqnarray} \label{NormZMAdS}
  \int\tilde U_0^2dz= \int(\cos(\frac{H}{\delta}z))^{(2q-3)\delta}e^{t_1
                \big[1-e^{\frac{4H^2t_2(3\delta+2)}{\delta}(\cos(\frac{H}{\delta}z))^{2\delta-2}}\big]}dz<\infty.
\end{eqnarray}
Then, it can be seen that this condition is satisfied with $(2q-3)\delta>1$. As parameter $\delta>1$, if $q<3/2$, the
zero mode cannot be normalized, while if $q>3/2$, it can.

For this AdS$_4$ brane case, the localization of the scalar fields has been studied in previous works
\cite{YXLiuJHEP1002,YXLiuPRD1184,guo2023scalar}. The localization of the scalar zero mode requires the
introduction of a coupling potential of the scalar field to itself and to the domain-wall-forming field
\cite{YXLiuJHEP1002,guo2023scalar}. Here, we focus on investigating the localization of the $U(1)$ gauge
vector fields and the KR fields within the AdS brane model (\ref{AdSbrane}).

\subsubsection{$U(1)$ gauge vector fields}  \label{AdSvector}

In this section, we will investigate the $q=1$ $U(1)$ gauge vector field within the AdS brane model
(\ref{AdSbrane}). Firstly for the vector zero mode, based on Eq. (\ref{zero mode}), we can express it as
\begin{eqnarray}  \label{ZMVecAdS}
  \rho_0(z) = \cos^{-\frac{\delta}{2}}\bigg(\frac{H}{\delta}z\bigg)
      \sqrt{e^{t_1\big[1-e^{\frac{4H^2t_2(3\delta+2)}{\delta}(\cos(\frac{H}{\delta}z))^{2\delta-2}}\big]}}.
\end{eqnarray}
With parameter $\delta>1$, we can see that this zero mode diverges to positive infinity as $z\rightarrow \pm z_b$.
Consequently, the vector zero mode cannot be localized on the AdS$_4$ brane.

\begin{figure} 
\begin{center}
\subfigure[$V_1(z)$.]{\label{FigEffPotVecAdS}
\includegraphics[width= 0.48\textwidth]{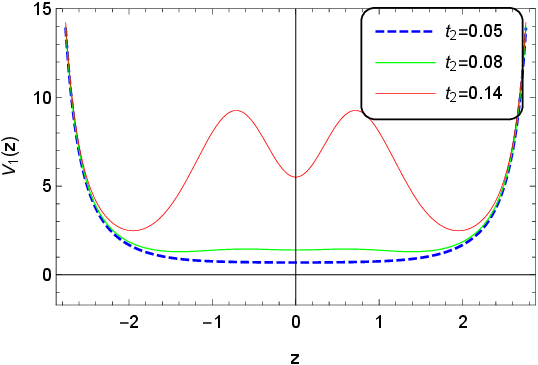}}
\subfigure[$\rho_0(z)$.]{\label{FigZMVecAdS}
\includegraphics[width= 0.48\textwidth]{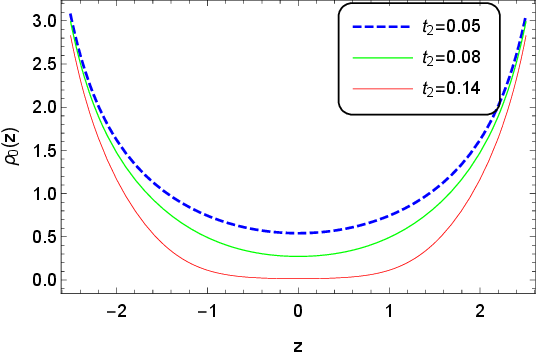}}
\end{center}\vskip -5mm
\caption{For the AdS$_4$ brane case, the effective potentials $V_1(z)$ and the zero mode $\rho_0(z)$
         of the $U(1)$ gauge vector fields. The parameters are set as $\delta=2,H=1, t_1=1$, and
         $t_2=0.05,0.08,0.14$.}
 \label{FigEffPotZMVecAdS}
\end{figure}

Then, for the massive vector modes, their localization is determined by the behaviors of the
effective potential (\ref{eff-potential}). As $z\rightarrow \pm z_b$, the coupling function
$F(R)$ tends to value $1$, and the coupling returns to the minimal coupling. Hence, the effective
potential exhibits the following asymptotic behavior as
$z\rightarrow \pm z_b$:
\begin{eqnarray}  \label{AsymEffPotVecAdS}
  V_1(z\rightarrow \pm z_b)&\rightarrow& \frac12A''+\frac14A'^2                              \nonumber   \\
                       &=& \frac{H^2}{4\delta}\big[-\delta+(\delta+2)\sec^2(\frac{H}{\delta}z)\big].
\end{eqnarray}
We can see that the effective potential diverges at the boundaries of the
extra dimension. Therefore, all massive KK modes will be localized on the thick AdS$_4$ brane. We plot
the effective potential $V_1(z)$ and the zero mode $\rho_0(z)$ in Fig. \ref{FigEffPotZMVecAdS}
for certain values of parameters. The effective potential is always positive, and diverges
to positive infinity as $z\rightarrow\pm z_b$. Furthermore, as parameter $t_2$ rises, the effective
potential increases at brane position, and then the lower localized massive modes will exhibit
larger masses.

\subsubsection{Kalb-Ramond fields}  \label{AdSKR}

Considering the KR fields, firstly for the zero mode, we can derive its expression in terms of Eqs. (\ref{AdSbrane})
and (\ref{FRAdS}):
\begin{eqnarray}  \label{ZMKRAdS}
  U_0(z) = \cos^{\frac{\delta}{2}}\bigg(\frac{H}{\delta}z\bigg)
      \sqrt{e^{t_1\big[1-e^{\frac{4H^2t_2(3\delta+2)}{\delta}(\cos(\frac{H}{\delta}z))^{2\delta+2}}\big]}}.
\end{eqnarray}
From this expression, it can be seen that this zero mode tends to zero at the boundaries of the
extra dimension. However, the region of the extra dimension becomes infinitely large in terms
of coordinate $y$. So, the normalization of this zero mode (\ref{ZMKRAdS}) requires $\delta>2$.

Then, for the massive KK modes, we discuss the behaviors of the effective potential (\ref{eff-potential}),
as the same. At the boundaries of the extra dimension, the coupling function $F(R)$ tends to
value $1$, so the effective potential becomes
\begin{eqnarray}  \label{AsymEffPotKRAdS}
  V_{\text{KR}}(z\rightarrow \pm z_b)&\rightarrow& -\frac12A''+\frac14A'^2                              \nonumber   \\
                       &=& \frac{H^2}{4\delta}\big[-\delta+(\delta-2)\sec^2\big(\frac{H}{\delta}z\big)\big].
\end{eqnarray}
For this expression, we can further obtain the following asymptotical behaviors for the effective
potential:
\begin{eqnarray}  \label{ClassAsymEffPotKRAdS}
  V_{\text{KR}}(z\rightarrow \pm z_b)&\rightarrow&
      \left\{
        \begin{array}{ll}
          +\infty  & \hspace{0.5cm}  \delta>2, \\
          -\frac{H^2}{4}  & \hspace{0.5cm} \delta=2, \\
          -\infty  & \hspace{0.5cm} 1<\delta<2.
        \end{array}
      \right
      .
\end{eqnarray}
It can be seen that if the parameter $\delta\in(1,2]$, the effective potential remains negative
at the boundaries of the extra dimension, and the zero mode can move far away from the brane.
Therefore, for the zero mode to be localized, the parameter $\delta$ must satisfy $\delta>2$.

\begin{figure} 
\begin{center}
\subfigure[$V_{\text{KR}}(z)$.]{\label{FigEffPotKRAdS}
\includegraphics[width= 0.48\textwidth]{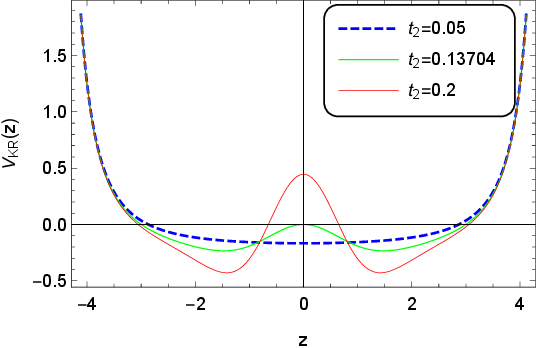}}
\subfigure[$U_0(z)$.]{\label{FigZMKRAdS}
\includegraphics[width= 0.48\textwidth]{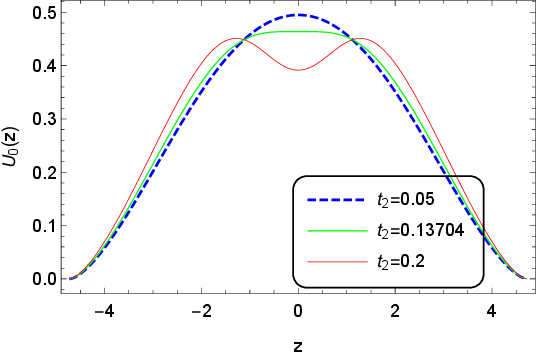}}
\end{center}\vskip -5mm
\caption{For the AdS$_4$ brane case, the effective potentials $V_{\text{KR}}(z)$ in (a), and the
         shapes of the zero mode $U_0(z)$ for the KR field in (b). The parameters are set as
         $\delta=3,H=1,t_1=0.05$, and $t_{2}=0.02,0.13704,0.2$.}
 \label{FigZMEffPotKRAdS}
\end{figure}

We describe the effective potential $V_{\text{KR}}(z)$ and the zero mode $U_0(z)$ visually in
Fig. \ref{FigZMEffPotKRAdS} with certain parameter values. The effective potential consists of
a series of infinitely deep wells, and the zero mode can be localized on the thick brane.

\begin{figure} 
\begin{center}
\includegraphics[width= 0.425\textwidth]{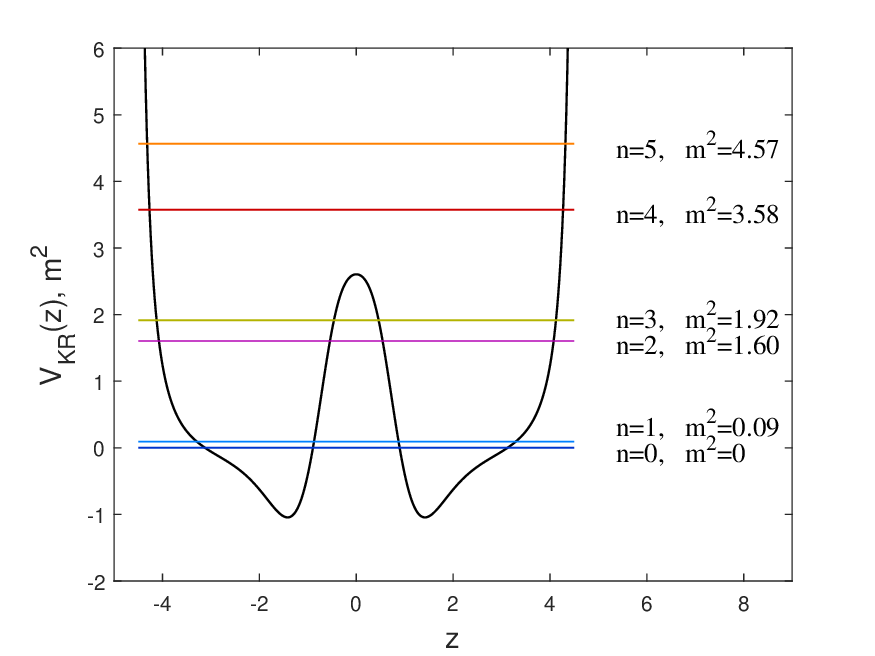}
\includegraphics[width= 0.56\textwidth]{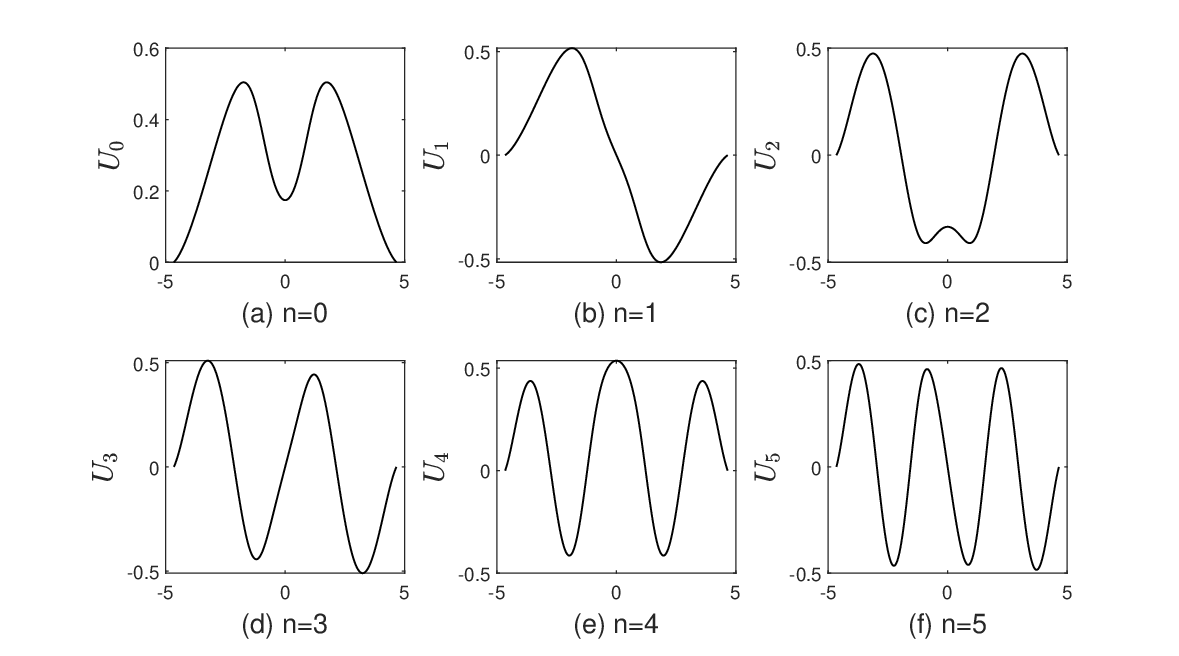}
\end{center}\vskip -5mm
\caption{For the AdS$_4$ brane case, the effective potentials $V_{\text{KR}}(z)$ and the mass spectra
         in the left. The zero mode $U_0(z)$ and the lower massive modes $U_n(z)$ in the right. The parameters
         are set as $\delta=3,H=1,t_1=0.05$, and $t_{2}=0.28$.}
 \label{FigEnerSpecKRAdS}
\end{figure}

Moreover, as parameter $t_2$ increases, the initially negative effective potential becomes positive,
with a barrier forming at the origin of the extra dimension. Consequently, the maximum of the zero
mode $U_0(z=0)$ transitions into a local minimum. The mass spectra and the lower localized modes
are depicted in Fig. \ref{FigEnerSpecKRAdS} using numerical methods. At the origin $z=0$, the height
of the potential barrier surpasses the energy level of the third massive mode $U_3$. This potential
well leads to the change in the extremum type of the even-parity modes $U_0$ and $U_2$, while
the odd-parity modes $U_1$ and $U_3$ are less affected. Additionally, for the KK modes with square
masses lower than the height of potential barrier, it is observed that their energy levels approach
in pairs.

\begin{figure} 
\begin{center}
\includegraphics[width= 0.48\textwidth]{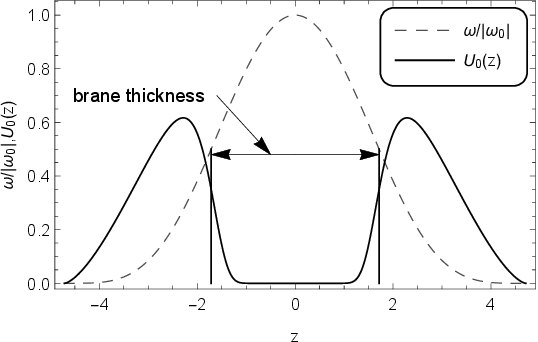}
\end{center}\vskip -5mm
\caption{For the AdS$_4$ brane case, the relative energy density $\omega/|\omega_0|$ of the
         AdS$_4$ brane model (\ref{AdSbrane}), and the zero mode $U_0(z)$ for the KR field.
         The parameters are set as $\delta=3,H=1,t_1=0.05$, and $t_{2}=0.5$.}
 \label{FigOriZMEffPotKRAdS}
\end{figure}

Then, if the parameter $t_2$ increases further, the potential barrier will be higher, and the zero mode
will be suppressed to zero at the origin. For this case, the zero mode $U_0(z)$ and the relative energy
density $\omega/|\omega_0|$ of the brane model (\ref{AdSbrane}) are shown in Fig. \ref{FigOriZMEffPotKRAdS}.
The zero mode of the KR field is localized on both sides of the brane.

For this AdS$_4$ brane case, as the bulk spacetime is asymptotically flat, the coupling function $F(R)$
tends to value $1$ at the boundaries of the extra dimension, and the coupling returns to the minimal
coupling. Consequently, considering the coupling mechanism will not change the localization results
of the KK modes for the KR fields. If $\delta>2$, the zero mode, as well as the massive ones, can always be
localized on the thick AdS$_4$ brane. However, these KK modes could exhibit rich structures and behaviors
at finite positions of the extra dimension.

\section{Conclusions}\label{Cons}

In this paper, we introduce a coupling mechanism between the kinetic term of the $q-$form field and the
background spacetime. Based on this scenario, there is a multiplier factor $F(R)$ in the action of
the $q-$form field. This function $F(R)$ only depends on the scalar curvature of the bulk. Then, we further
investigate the localization of concrete $q-$form fields with three kinds of brane models.

In the case of the $\mathcal{M}_4$ brane, the function $F(R)$ is characterized by two parameters, $t_1$
and $t_2$. The parameter $t_2$ determines the shape of the effective potential of KK modes to be volcanic,
P\"{o}schl-Teller, or oscillator harmonic. Furthermore, the zero modes of various $q-$form fields
can be localized on the thick brane, and the massive ones can be localized, or quasi-localized on the
brane. Especially for the scalar fields, if $t_2<0$, a potential well forms with a positive lower
boundary at the origin of the extra dimension. Then, the scalar zero mode could be localized
on both sides of the brane, and the massive KK modes can exist as resonant states at the brane position.

In the case of the dS$_4$ brane, the localization of the $q=1$ $U(1)$ gauge vector field and the
$q=2$ KR field is analyzed. It is found that the vector zero mode can be localized on the brane,
while the zero mode of the KR field cannot. Moreover for the $U(1)$ gauge vector field, with larger
values of parameters in function $F(R)$, a finitely deep potential well exists with a positive lower
boundary at the origin of extra dimension. Owing to this configuration, the vector zero mode can
be localized on both sides of the brane, while the massive vector modes could be quasi-localized
at the brane position.

In the case of the AdS$_4$ brane, we present the localization results of the $q=1$ $U(1)$ gauge
vector field and the $q=2$ KR field. For the $U(1)$ gauge vector field, the zero mode cannot be
localized on the brane, while all massive modes can be localized. In contrast, for the KR field,
both the zero mode and all massive modes can be localized on the brane. Additionally for the
latter, as the parameter in the function $F(R)$ increases, the effective potential becomes
positive at the origin of the extra dimension, and then the zero mode and low-lying massive
modes could be localized on both sides of the brane.

For the coupling mechanism considered here, it describes a physically motivated, and rational
relation between the kinetic term of the $q-$form field and the background spacetime. In this
framework, if the bulk is regular and asymptotically bent, various $q-$form fields can be
localized on the brane. On the other hand, if the bulk is asymptotically flat, the function
$F(R)$ approaches value $1$ when far away from the brane, and the coupling returns to the
minimal coupling. In this case, the localization of the KK modes for the $q-$form fields
remains unchanged. However, these KK modes will behave diversely at finite positions of the
extra dimension.

\section*{Acknowledgments}

This work is supported by the National Natural
Science Foundation of China (Grants No. 11305119), the Natural Science Basic Research Plan in Shaanxi
Province of China (Program No. 2020JM-198), the Natural Science Foundation of Shaanxi Province
(No. 2022JQ-037), the Fundamental Research Funds for the Central Universities (Grants No. JB170502),
and the 111 Project (B17035).


\end{document}